\documentclass[reprint]{revtex4-1}
\usepackage{graphics}
\usepackage{graphicx}
\usepackage{amsmath}
\graphicspath{{./}{./Figures/}}

\begin{document}

\title{
Anisotropic avalanches and critical depinning of three-dimensional magnetic domain walls
} 

\author{Joel T. Clemmer}
\affiliation{Department of Physics and Astronomy, Johns Hopkins University, Baltimore, Maryland 21218, USA}
\author{Mark O. Robbins}
\affiliation{Department of Physics and Astronomy, Johns Hopkins University, Baltimore, Maryland 21218, USA}
\date{\today}

\begin{abstract}
	Simulations with more than $10^{12}$ spins are used to study the motion of a domain wall driven through a three-dimensional random-field Ising magnet (RFIM) by an external field $H$.
The interface advances in a series of avalanches whose size diverges at a critical external field $H_c$.
Finite-size scaling is applied to determine critical exponents and test scaling relations.
Growth is intrinsically anisotropic with the
height of an avalanche normal to the interface $\ell_\perp$ scaling as the width along the interface $\ell_\|$ to a power $\chi=0.85 \pm 0.01$.
The total interface roughness is consistent with self-affine scaling with
a roughness exponent $\zeta \approx \chi$ that is much larger
than values found previously for the RFIM and related models that explicitly break orientational symmetry by requiring the interface to be single-valued.
Because the RFIM maintains orientational symmetry, the interface develops overhangs that may surround unfavorable regions to create uninvaded bubbles.
Overhangs complicate measures of the roughness exponent but decrease in importance with increasing system size.
\end{abstract}

\maketitle

\section{Introduction}

Many disordered systems exhibit critical behavior when they are driven slowly \cite{Sethna2001, Fisher1998, Kardar1997}.
Evolution occurs through a scale-free sequence of bursts or avalanches that have a power-law distribution of sizes.
Notable examples of such avalanches include earthquakes \cite{Gutenberg1954}, sandpile avalanches \cite{Bak1987}, Barkhausen noise in magnetization \cite{Cote1991, Perkovic1995}, and the jerky advance of an elastic interface through a medium with quenched disorder \cite{Fisher1983}.
Here we focus on the latter case, which is important in magnetic domain wall motion \cite{Urbach1995, Ji1991, Ji1992, Koiller1992, Koiller1992a, Nolle1996, Koiller2000, Koiller2010, Roters1999}, fluid invasion in porous media \cite{Cieplak1988, Martys1991, Martys1991a,Stokes1988}, contact line motion \cite{Stokes1988, Stokes1990, Joanny1990, Ertas1994, Duemmer2007}, and the propagation of crack fronts \cite{Gao1989, Ramanathan1997, Maloy2006}.

The onset of athermal motion of a driven interface is called a depinning transition and occurs at a critical driving force $F_c$.
As $F$ increases towards $F_c$ the interface advances between stable states in a sequence of avalanches.
The size of avalanches grows as $F$ approaches $F_c$ and can be related
to a diverging correlation length.
For $F>F_c$ the interface is never stable, and avalanches are associated with fluctuations in the interface velocity. As $F$ increases, these fluctuations become smaller.
The value of $F_c$ is determined by a competition between the disorder and the elastic cost of deforming the interface.
Different universality classes have been identified, depending on whether
disorder is large or small \cite{Cieplak1988, Martys1991, Martys1991a, Ji1991, Ji1992,  Koiller1992, Koiller1992a, Nolle1996, Koiller2000, Koiller2010} and whether
elastic interactions are local or have a long, power-law tail \cite{Ramanathan1997, Duemmer2007}.

A magnetic domain wall or fluid interface can have any orientation in a $d$ dimensional system and the driving force always favors advance perpendicular to the local orientation.
At high disorder, the advancing interface becomes self-similar with a fractal dimension related to percolation \cite{Ji1991, Ji1992, Koiller1992, Koiller1992a, Nolle1996, Koiller2000, Koiller2010, Cieplak1988, Martys1991, Martys1991a}.
At low disorder, elastic interactions are able to spontaneously break symmetry and enforce an average interface orientation
\cite{Ji1991, Ji1992, Koiller1992, Koiller1992a, Nolle1996, Koiller2000, Koiller2010, Cieplak1988, Martys1991, Martys1991a}.
The interface becomes self-affine, and fluctuations in height along the average surface normal rise as $\ell^\zeta$ where $\zeta < 1$ is the roughness exponent and $\ell$ is the displacement in the $d-1$ dimensions along the interface.

Most models of interface motion focus on the self-affine regime and begin with the assumption that the height is a single-valued function \cite{Narayan1993, Nattermann1992, Amaral1994, Kardar1986, Chauve2001, Rosso2003, Rosso2009, LeDoussal2009}.
While this simplifies the application of analytical methods, it explicitly
breaks the spatial symmetry of the physical system and may thus change the universality class.
The above models also use an approximation for the elastic energy that is valid only when derivatives of the height are much less than unity.
These assumptions may not be self-consistent because the regions of extreme disorder, which are important to pinning, also create large forces and therefore large surface slope and curvature \cite{Coppersmith1990, Cao2018}.
Moreover, motion can be stopped at any field by a single unflippable spin or uninvadable pore.
Such extreme regions need not stop a fully $d$-dimensional interface.
A multi-valued interface can have overhangs that advance around regions of strong disorder and merge to create enclosed bubbles that are left behind the advancing interface.
This process is clearly observed in advancing fluid interfaces \cite{Krummel2013}.

In this paper we examine the critical depinning transition in a model that does not impose an interface orientation, the $d = 3$ random field Ising model (RFIM) \cite{Ji1992, Nolle1996, Koiller2000, Koiller2010}.
Simulations with more than $10^{12}$ spins are analyzed using finite-size scaling, and scaling relations between exponents are derived and tested.
While the domain wall between up and down spins is not single-valued, growth is strongly anisotropic.
The correlation lengths along and perpendicular to the interface diverge near the critical point with different exponents $\nu_\|$ and $\nu_\perp$, respectively.
Individual avalanches show the same growing anisotropy, with the height scaling as width to the power $\chi=\nu_\perp/\nu_\| = 0.85\pm 0.01$.
The anisotropy is also consistent with scaling relations for the distribution of avalanche volumes and lengths, and for the maximum volume and lengths.
The scaling of the total root mean square (rms) interface roughness is consistent with $\zeta=\chi$. The power law describing changes in roughness with separation along the interface appears to approach $\chi$ as $L$ increases near the critical field.

These results are quite different from earlier work on the RFIM.
Calculated
exponents were consistent with scaling relations that assumed $\chi =1$ \cite{Martys1991a}, but used systems with linear dimensions more than 40 times smaller \cite{Ji1992} for which we show finite-size effects are significant.
Later work  \cite{Nolle1996} used systems up to four times larger and found $\chi =0.9 \pm 0.1$, which is still consistent with unity.
All earlier work \cite{Ji1992, Koiller2000, Koiller2010} concluded that the roughness exponent $\zeta$ was consistent with the mean-field value of $2/3$ and less than $\chi$.

Our results show that the ratio of overhang height to interface width decreases with increasing system size. This suggests that the RFIM might be in the same universality class as models that assume the interface is single-valued. 
The results are compared to studies of the quenched Edwards-Wilkinson (QEW) equation, a single-valued interface model that is often used for domain wall motion \cite{Edwards1982, Chauve2001, Rosso2009, LeDoussal2009}.
Some exponents, such as the power law describing the distribution of avalanche volumes, are nearly the same in both models \cite{Rosso2009,Ji1992}.
However, the anisotropy is quite different.

In Sec. II we describe the implementation of the RFIM model and different growth protocols used. Results are presented in Sec. III. In Sec. IIIA, the critical field and correlation length exponent are first identified using the fraction of avalanches which span the system. Next the divergence of avalanches as the system approaches the critical field and the distribution of avalanches at the critical field are calculated in Secs. IIIB and IIIC, respectively. We next look at the morphology of avalanches in Sec. IIID and the scaling of spanning avalanches in Sec. IIIE. In Secs. IIIF and IIIG we study the scaling of the interface morphology. Finally, the contribution of overhangs to the total interfacial width is discussed in Sec. IIIH. In Sec. IV we summarize our results and compare to past work.

\section{Methods}

We simulate athermal motion of a domain wall in the RFIM on a cubic lattice in $d = 3$ \cite{Ji1992,Koiller2000}. The Hamiltonian of the system is given by
\begin{equation}
\mathcal{H} = \sum_{<i,j>} s_i s_j - \sum_i (\eta_i + H) s_i \ \ ,
\end{equation}
where $s_i = \pm 1$ is the state of the $i^\mathrm{th}$ spin, $H$ is the external magnetic field, and $\eta_i$ is the local random field. Interactions extend only to nearest neighbors, and the coupling strength is defined as the unit of energy. The nearest-neighbor spacing is defined as the unit of length. The random local field is taken to be Gaussian distributed with a mean of zero and a standard deviation of $\Delta$. 

Previous work has determined that there exists a critical value of the noise $\Delta_c \sim 2.5$ separating two universality classes \cite{Koiller2000}. In the limit of $\Delta > \Delta_c$, fluctuations in noise dominate the Hamiltonian such that interactions become irrelevant. Therefore, the local orientation of the interface does not significantly favor a direction of growth, and the problem reduces to invasion percolation \cite{Wilkinson1983, Ji1992, Koiller2000}.  The invaded volume has a self-similar hull described by percolation theory \cite{Adler1990}. In the limit of small noise, $\Delta < \Delta_c$, interactions lead to more compact, cooperative growth producing a self-affine interface \cite{Ji1992,Koiller2000}.
We have studied systems at a range of $\Delta$ and verified the transition from isotropic growth above $\Delta_c$ to anisotropic growth below $\Delta_c$.
Exponents for several $\Delta$ below $\Delta_c$ were consistent, and we
focus on results for $\Delta=1.7$ below.

Interfaces are grown with fixed boundary conditions along the direction of growth and periodic boundary conditions perpendicular to growth. The upper and lower boundaries consist of layers of down and up states, respectively, necessitating the presence of a domain wall within the bulk. In the periodic directions, the system has a width of $L_x = L_y = L$ while the height of the box along the direction of growth is typically set to $L_z = 2L$. A larger vertical dimension helps ensure the upper boundary condition does not interfere with growth for most simulation runs.

Systems are initialized with all spins in the down state except for the bottom layer, creating an initially flat domain wall.
Spins are allowed to flip up only if they lie on the interface i.e., if at least one of their neighbors is up.
This requirement is motivated by models with a conservation law such as fluid invasion where fluid must flow along a connected path to new regions \cite{Wilkinson1983, Cieplak1988, Martys1991, Martys1991a, Ji1991}.
As down spins adjacent to the interface flip up, the set of spins defining the interface evolves. These growth rules ensure
that there is a single domain wall separating the unflipped region at large $z$ from flipped spins at low $z$, as is usually assumed in scaling theories of interface motion through a disordered medium \cite{Nattermann1992,Narayan1993, Amaral1994, Kardar1986, Chauve2001, Rosso2003, Rosso2009, LeDoussal2009}.
In contrast, studies of Barkhausen noise in hysteresis loops of the RFIM allow disconnected spins to flip and this changes things like the critical disorder $\Delta_c$ \cite{Sethna1993,Perkovic1995,Perkovic1999}. 

The RFIM considered here has only cubic symmetry, but past studies show that scaling of interface growth is isotropic in both the self-affine and self-similar regimes \cite{Amaral1994, Amaral1995b, Koiller2010}. Planar growth along facets of the lattice occurs only for a bounded distribution of random fields at very weak disorder \cite{Koiller2010,Roters1999}. 
This is in sharp contrast to models that explicitly break symmetry by assuming the interface is a single-valued function of height 
\cite{Edwards1982, Chauve2001, Rosso2009, LeDoussal2009,Amaral1994,Amaral1995b}.
Some $2+1$ dimensional models even have direction-dependent critical fields and other anisotropic properties \cite{Amaral1994, Amaral1995b}.  
Given the established isotropy of growth in our model we consider the simplest case where the sides of the box are aligned with the nearest-neighbor directions and the initial interface has a (001) orientation.

\begin{figure*}
	\includegraphics[width=\textwidth]{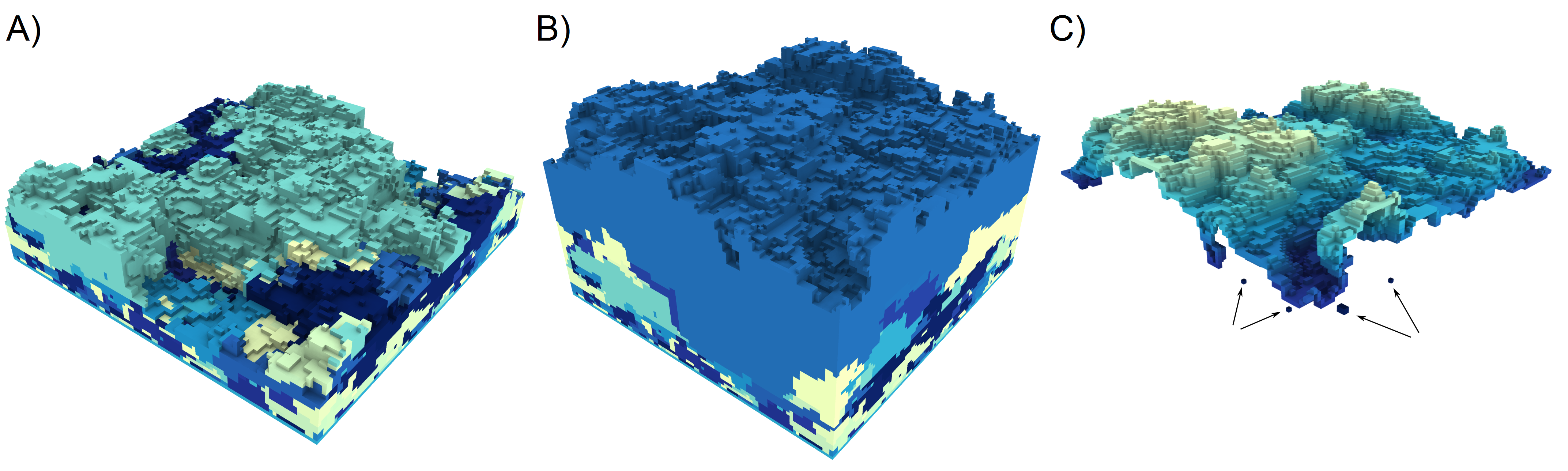}
	\caption{Flipped spins in a sample simulation of size $L = 100$ are shown for different stages of growth corresponding to (a) $H$ near $1.44579$ and (b) $H$ near $1.45853$. Contiguous spins are grouped by their associated avalanche and colored accordingly. Along the cross section one can see more examples of small avalanches at low heights which grew at smaller values of the external field $H$. At larger heights, growth occurred at a higher value of $H$ and larger avalanches are visible. The final avalanche (teal) in panel (a) is an example of a semispanning avalanche that wraps across a periodic boundary condition and percolates. The final avalanche (blue) in panel (b) is an example of a fully spanning avalanche that has a footprint of $L^2$ and advances the entire interface. The set of spins which could potentially flip in response to an increase in $H$ in panel (b) are rendered in panel (c). These unflipped spins can either be on the external interface or contained inside a bubble. Four bubbles consisting of either six unflipped spins or a single unflipped spin are indicated by arrows. The remaining visible region constitutes the external interface. Note that it is not a single-valued function of height, and large overhanging regions are visible at the bottom left and right cross sections. Particles are colored by height for improved visibility.}
	\label{fig:Method0}
\end{figure*}

Growth occurs athermally through single spin-flip dynamics. The external magnetic field is initialized to the lowest value that will excite a single spin on the interface to flip up. The stability of neighboring down spins is checked, and they are flipped up if this lowers the global energy. This procedure can lead to a chain reaction and is repeated until all spins are stable along the interface. The ``no-passing rule" \cite{Middleton1992, Middleton1993} guarantees that the resulting interface is independent of the algorithmic order in which spins are flipped. The magnetic field is then increased to flip the least stable remaining spin, and the process is repeated until either the interface reaches the upper boundary or the field is well above the critical point. Fewer than 2\% of systems of size $L = 1600$ hit the boundary at a height of $3200$ before reaching the critical field. This growth algorithm produces invaded volumes such as the examples in Fig. \ref{fig:Method0} rendered using the Open Visualization Tool (OVITO) \cite{Stukowski2010}.

Each time the external field is incremented, the resulting cluster of flipped spins is recorded and grouped as a single avalanche. The volume and linear dimensions of each avalanche are calculated. The volume is simply the number of spins flipped, and the linear dimensions are calculated based on measures of the width and height defined in Sec. IIID. Because the interactions are short range, all spins in a cluster are connected. Some avalanches can have a length of $L$ or larger in the direction perpendicular to growth due to the periodic boundary conditions. If these avalanches percolate, colliding with a periodic image of themselves, specifying their lateral size is ambiguous. We will refer to these avalanches as spanning avalanches, and two examples are seen in Figs. \ref{fig:Method0}(a) and \ref{fig:Method0}(b). Due to this ambiguity and changes in scaling discussed in Sec. IIIE, we exclude spanning avalanches from most analyses unless otherwise mentioned.  Avalanches which are truncated by reaching the upper boundary are always excluded because their growth is artificially restricted.

We further divide spanning avalanches into two classes: semispanning and fully spanning avalanches. We define the footprint of an avalanche as the total area of all flipped spins projected onto the $x$-$y$ plane. The footprint of any avalanche is contained in the interval $[1, L^2]$ by definition. We define semispanning avalanches as percolating events that have a footprint less than $L^2$. A specific example is shown in Fig. \ref{fig:Method0}(a). Fully spanning avalanches are percolating events that have a footprint equal to $L^2$ such as the final avalanche seen in Fig. \ref{fig:Method0}(b). The differences between these two classes of spanning avalanches are discussed in Sec. IIIE. 

Any unflipped down spin with a neighbor in the flipped up state is a potential site for an avalanche. However, as seen in Fig. \ref{fig:Method0}(c), these spins can be sorted into two topologically distinct regions: the external interface and bubbles. The external interface consists of spins that are connected to the upper boundary of the cell by an unbroken chain of unflipped spins. This interface delimits the extent of propagation. Alternatively, certain spins with a strong pinning force may become surrounded by the domain wall and enclosed in a bubble. While avalanches could still grow in bubbles, they would be heavily constrained by the geometry of the bubble and would not contribute to the structure of the external interface. Therefore they are excluded from all analysis in this paper. This rule is analogous to the problem of incompressible fluid invasion where growth within bubbles is not allowed \cite{Wilkinson1983, Cieplak1988}.
The average fraction of volume behind the external interface that is in bubbles is quite small and nearly independent of $H$ and $L$.
The fraction of bubbles does depend on disorder, dropping from less than 0.02\% for $\Delta=2.1$ to below 0.001\% for $\Delta=1.7$. Since these fractions are low and avalanches inside bubbles are small, excluding avalanches in bubbles has little impact on the avalanche statistics, especially for the large sizes that dominate the critical behavior.

	In addition to the growth protocol described above, a second protocol was also implemented. In this method, the external field $H$ is set at a fixed value, and unstable spins are continually flipped until the interface is stable. For each ensemble, stable interfaces were found for a set of increasing $H$. The values of $H$ were chosen to thoroughly sample the approach to the critical field $H_c$, which is measured in Sec. IIIA. In this protocol, we do not resolve individual avalanches. This allows for efficient parallelization of the code allowing simulation of larger system sizes. As referenced before, the no-passing rule \cite{Middleton1992,Middleton1993} guarantees that the resulting interface does not depend on the parallelization scheme. Using the primary protocol and tracking individual avalanche growth we simulate systems up to a size of $L = 3200$. With the alternate protocol we reached system sizes of $L = 25600$, flipping more than $10^{12}$ spins. At all system sizes, many simulations were run with different realizations of disorder, and results were averaged.

\section{Results and Discussion}

\subsection{Determining the critical field}

As the external field is increased, the domain wall advances through a sequence of avalanches. The size of the largest avalanche increases with external field, indicating a growing correlation length. The critical field $H_c$ is defined as the field where the correlation length diverges and interfaces in an infinite system will depin and advance indefinitely. In a finite-size system the depinning transition is broadened. There is a range of $H$ where the correlation length is comparable to the system size $L$. In this range, interfaces in some systems will remain pinned while others will advance to the top. In this section we use finite-size scaling methods to determine $H_c$ and the scaling of the in-plane correlation length $\xi_\|$ from simulations with different $L$. 

For a self-affine system, correlations may be different for motion along and perpendicular to the interface. We define a correlation length along the interface as $\xi_\|$ and a correlation length in the direction of growth as $\xi_\perp$. Both are expected to diverge at the critical field in an infinite system with exponents $\nu_\|$ and $\nu_\perp$, respectively:
\begin{align}
\begin{split}
\xi_\| &\sim |H_c-H|^{-\nu_\|}\\ 
\xi_\perp &\sim |H_c-H|^{-\nu_\perp}.
\label{eqn:xis}
\end{split}
\end{align}
We define $\chi = \nu_\perp/\nu_\|$ such that $\xi_\perp \sim \xi_\|^\chi$.

The total volume invaded over an interval of external field is defined as the number of spins that become unstable and flip. For a finite system, a fraction $F_s$ of these flipped spins will be part of system-spanning avalanches, while the rest are in smaller avalanches. At very low fields where $\xi_\| \ll L$, no avalanches will span the system and $F_s = 0$.  At very large fields, $H > H_c$, $F_s \rightarrow 1$ as the system becomes depinned at all system sizes and the largest, spanning avalanches dominate the increase in volume. 

\begin{figure}
	\includegraphics[width=0.45\textwidth]{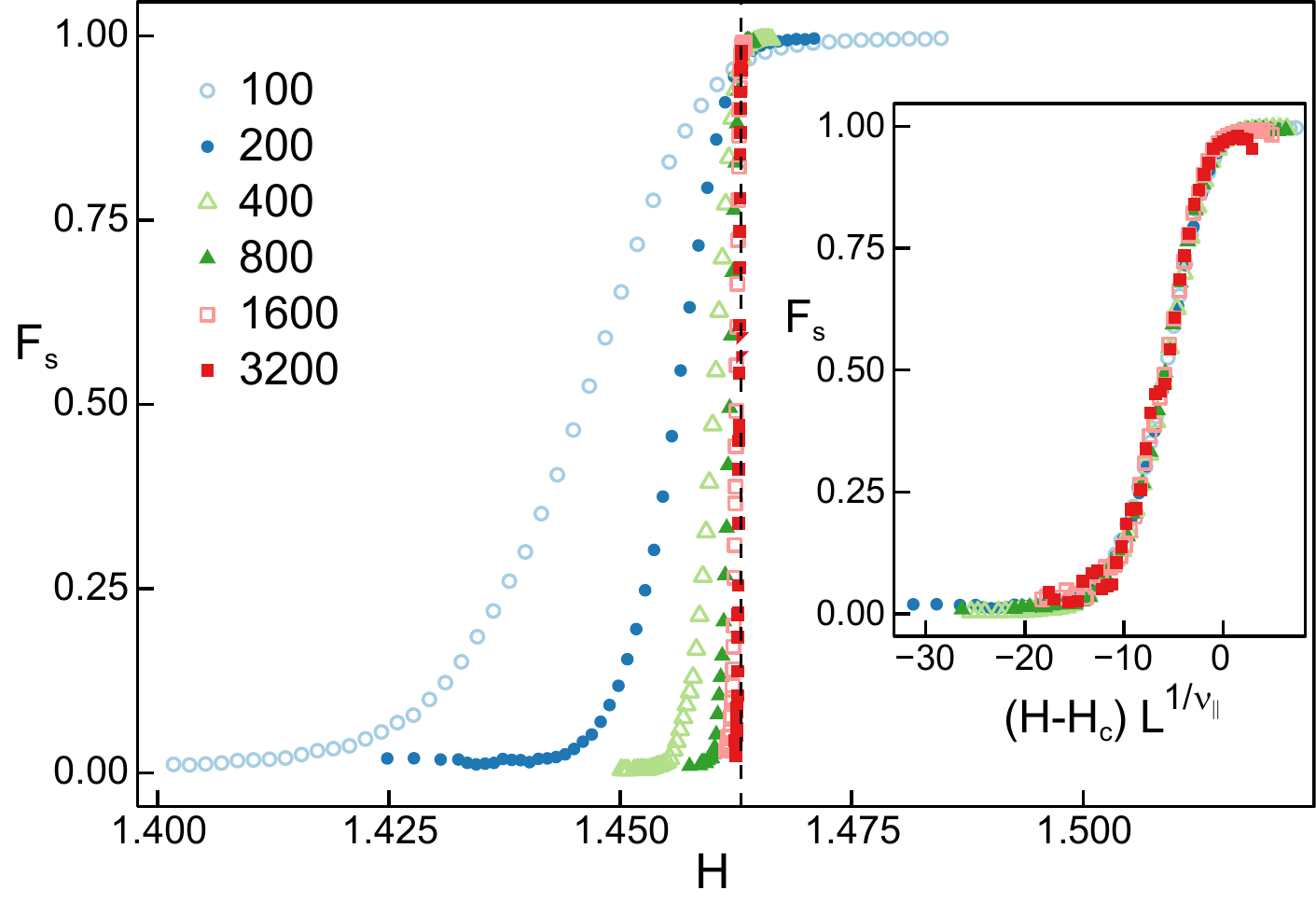}			
	\caption{The fraction of volume invaded due to system spanning avalanches over a small interval of $H$ is calculated for the values of L indicated in the legend. A dashed vertical line indicates $H_c = 1.46305$. The inset shows the collapsed data using the finite-size scaling procedure described in Eq. \eqref{eqn:fs} with a value of $H_c = 1.46305$ and $\nu_\| = 0.79$.}
	\label{fig:V0}
\end{figure}

Figure \ref{fig:V0} shows the change in $F_s$ with $H$ for different system sizes. For each $L$, the size of increments in $H$ was chosen to be small enough to resolve the transition but large enough to reduce noise. After calculating $F_s$ for each interval, the curves were further smoothed by applying a rolling average across all sets of three adjacent intervals.  At fields above $H_c$, many systems have already reached the top of the box and stopped evolving. We therefore discarded poorly sampled data points at large values of $H - H_c$. The transition from growth by finite avalanches to spanning avalanches sharpens as $L$ increases. Using a simulation cell of height $2L$ ensured that $F_s$ was not significantly affected by finite system height for fields near and below $H_c$.

In finite-size scaling theory one assumes that the only important length scales in the system are the correlation lengths, $\xi_\|$ and $\xi_\perp$, and the system size, $L$. Finite-size effects are expected when the largest correlation length approaches $L$. The simulation cell is taller than it is wide, and we find $\xi_\perp < \xi_\|$, so $\xi_\|$ dominates the finite-size effects. Functions like $F_s$ then depend on the dimensionless scaling variable $L/\xi_\|$. Using Eq. \eqref{eqn:xis}, $F_s$ can be expressed in terms of the field as
\begin{equation}
F_s \sim f\left((H-H_c) L^{1/\nu_\|} \right) \ \ ,
\label{eqn:fs}
\end{equation}
where the scaling function $f$ should be independent of $L$. Given the limiting behavior of $F_s$, $f(x)$ must approach zero for $x \ll -1$ and one for $x \gg 1$. Note that Eq. \eqref{eqn:fs} gives $F_s = f(0)$ for all $L$ at $H = H_c$. Therefore the critical field must correspond to the location where all curves cross in Fig. \ref{fig:V0}. This intersection occurs at a value of $H_c \sim 1.46304 \pm 0.00003$. Here and below, the error bars do not represent a standard deviation, but indicate the maximum range over which data collapse within statistical fluctuations. Koiller and Robbins had previously found $H_c$ for various values of $\Delta$ in this system \cite{Koiller2000}. Although the value of $H_c$ was not explicitly determined for $\Delta = 1.7$, our result is consistent with interpolations of their data from nearby values of $\Delta$.

Equation \eqref{eqn:fs} also implies that all curves should collapse when plotted against $(H-H_c)L^{1/\nu_\|}$ for the correct value of $\nu_\|$. For all scaling collapses in the following plots, we choose to use a common value of $H_c = 1.46305$ based on consideration of the above estimate of $H_c$ and the scaling of other system properties discussed later in the paper. Table \ref{table:1} contains the values of all scaling exponents used in the plots and the estimated uncertainties.

The inset of Fig. \ref{fig:V0} shows a successful collapse of $F_s$ with a value of $\nu_\| = 0.79$. Based on the sensitivity of the collapse to changes in $\nu_\|$, the data are consistent with $\nu_\| = 0.79 \pm 0.02$. This value is close to prior estimates of $\nu_\| = 0.75 \pm 0.02$ \cite{Koiller2000} and $0.75 \pm 0.05$ \cite{Ji1992} in the RFIM. References \cite{Ji1992} and \cite{Koiller2000} used scaling approaches that assume $\chi = 1$, which may have impacted the reported values. The RFIM with a uniform instead of Gaussian distribution of random fields is expected to be in the same universality class and past simulations found $\nu_\| = 0.77(4)$ \cite{Roters1999}. For the QEW model of interface growth, the mean-field value of $\nu$ is found to be $3/4$ \cite{Nattermann1992}. Arguments in Ref. \cite{Narayan1993} suggest 3/4 is a lower bound on the actual exponent.
Epsilon expansions \cite{Chauve2001} give 0.67 and 0.77 to first and second order, which suggests that $\nu$ could be slightly above the mean-field value.

The maximum distance a depinning avalanche can advance the interface is set by the box height. One might wonder whether this artificial threshold could affect the scaling of $F_s$. As an alternative measure, we considered the footprint of an avalanche, the projected area in the $x$-$y$ plane of all spins flipped by an avalanche. This measure is independent of how far an avalanche propagates in the $\hat{z}$ direction. Over an interval of $H$, avalanches will cumulatively advance the interface over a region equal to the sum of their footprints. Note that some avalanches may overlap such that certain regions may advance more than once. In analogy to $F_s$, one can then define the fraction of the area advanced by spanning avalanches, $F_a$. We find $F_a$ scales in the same manner as $F_s$ with consistent estimates of $H_c$ and $\nu_\|$. This verifies that the results of $F_s$ are not affected by alternative scaling behavior of spanning avalanches.

\begin{figure}
	\includegraphics[width=0.45\textwidth]{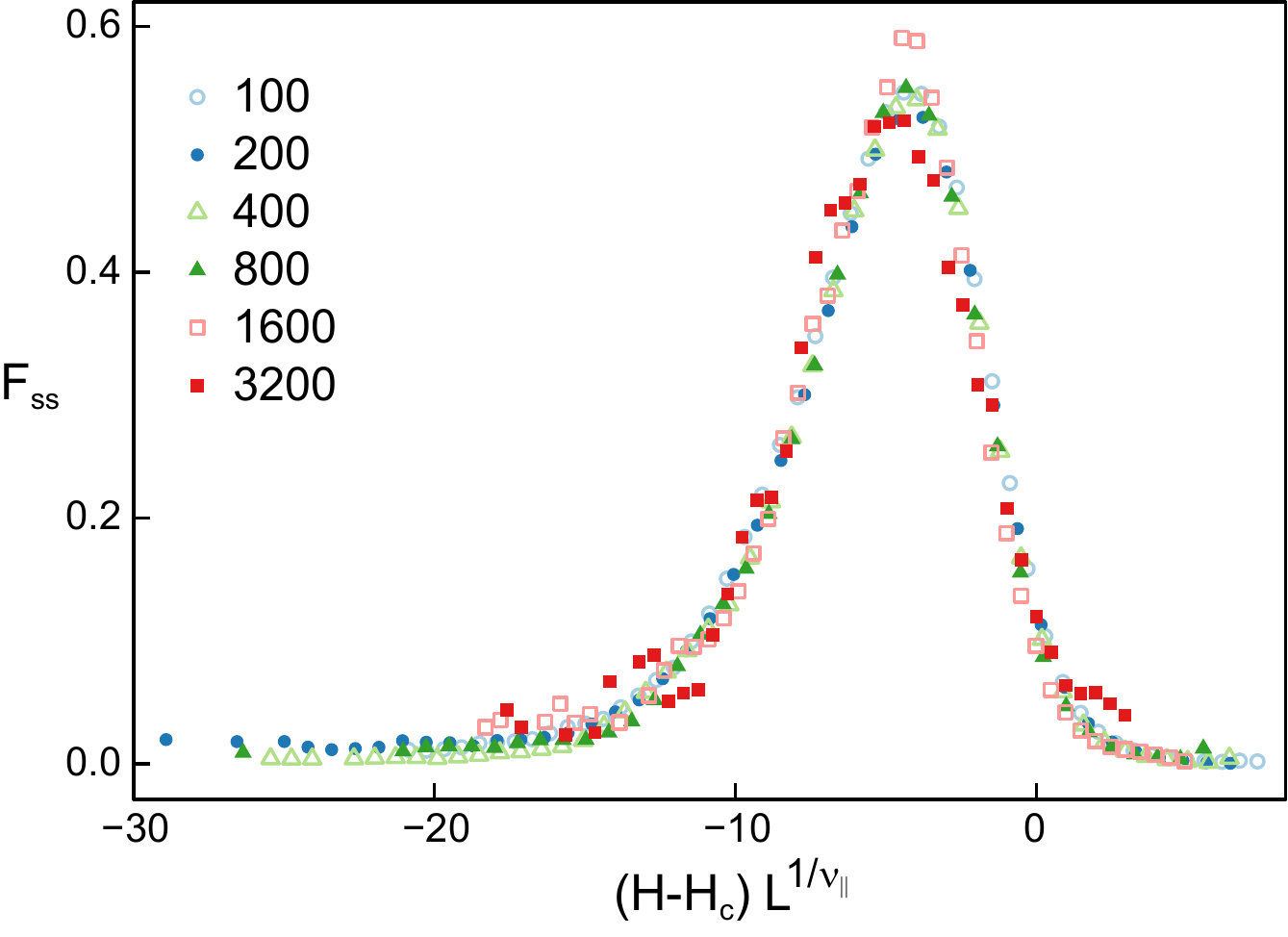}				
	\caption{The fraction of volume invaded due to semispanning avalanches over a small interval of $H$ is calculated for the values of $L$ indicated in the legend and scaled according to a finite-size scaling procedure similar to Eq. \eqref{eqn:fs}. The collapse uses values of $H_c = 1.46305$ and $\nu_\| = 0.79$.}
	\label{fig:V0.5}
\end{figure}

Another useful measure is $F_{ss}$, the fraction of growth in semispanning avalanches. Figure \ref{fig:V0.5} shows that $F_{ss}$ obeys a scaling relation like \eqref{eqn:fs} with the same $H_c$ and $\nu_\|$ but a different scaling function $f_{ss}(x)$. For each $L$, $F_{ss}$ rises from zero at small $H$ to a maximum below $H_c$ and then drops as fully spanning avalanches begin to dominate growth. From Figs. \ref{fig:V0} and \ref{fig:V0.5}, we see that semispanning and fully spanning avalanches begin to be important when $H_c -H$ is smaller than about $10 L^{1/\nu_\|}$ and $5 L^{1/\nu_\|}$, respectively. This is useful in estimating the region where $\xi_\| < L$.

Note that $f(0)$ has a value of about 0.97 that is very close to unity. This implies that almost all the incremental growth near $H_c$ is due to spanning avalanches. Spanning avalanches also make up roughly 75\% of the cumulative invaded volume from the initial flat interface to $H_c$. The importance of large avalanches is related to the power-law distribution of avalanche sizes that we discuss in the next two sections.

\subsection{Divergence of avalanches near $H_c$}

As noted above, spanning avalanches are more related to depinning above $H_c$ than the approach to $H_c$ from below. In addition, their height is bounded only by the arbitrary height of the simulation box. In contrast, the vertical growth of nonspanning avalanches is naturally correlated to their lateral extent. Thus in the next three sections we focus on nonspanning avalanches, providing a discussion of spanning avalanches in Sec. IIIE. Nonspanning avalanches that grow close to $H_c$, after the appearance of spanning avalanches, are included because they exhibit the same scaling as avalanches grown prior to the first spanning avalanche. Including them improves statistics without changing the exponents.

We define a normalized probability distribution of nonspanning avalanche volumes $S$, $P(S,H,L)$, which depends on both the current value of the field $H$ and the size of the system $L$. At the critical point, the distribution of avalanches is expected to decay as a power law with an exponent $\tau$, $P(S, H_c, \infty) \sim S^{-\tau}$. Away from the critical point the power law will extend to a maximum volume, $S_\mathrm{max}$, that reflects the influence of a limiting length scale $\ell$. In general this will be the smaller of the system size $L$ and the correlation length $\xi_\|$. The maximum volume will scale as power of this length, $\ell^\alpha$, where $\alpha$ is another critical exponent.

Having defined the behavior of the distribution, we can determine how statistical moments of avalanches depend on $S_\mathrm{max}$. The $m^\mathrm{th}$ moment of the avalanche volume is calculated by integrating the distribution up to the maximum avalanche cutoff $S_\mathrm{max}$:
\begin{align}
\langle S^m \rangle & = \int P(S, H, L) S^m dS\\
\langle S^m \rangle & \sim \int^{S_\mathrm{max}} S^{m-\tau} dS
\label{eqn:ave}
\end{align}
For values of $m > \tau - 1$, this integral is dominated by the largest avalanches and scales as:
\begin{equation}
\langle S^m \rangle \sim S_\mathrm{max}^{m-\tau + 1}.
\label{eqn:ave2}
\end{equation}
Alternatively, if $m < \tau - 1$, the integral is dominated by the smallest avalanches and will not diverge as a power of $S_\mathrm{max}$ but instead saturate. As shown next, the integral diverges for $m=1$, but not for $m=0$. This implies that $1 < \tau < 2$ and that $P(S,H,L)$ is independent of $H$ and $L$ for small $S$.

We will focus on the average size $\langle S \rangle$ ($m = 1$) as the lowest moment that gives information about $S_\mathrm{max}$. We define the variable $\Delta  H \equiv H_c - H$ as the distance to the critical field from below. To study the variation of $\langle S \rangle$ with $\Delta H$, $S$ is averaged over all nonspannning avalanches that nucleated in an interval of field. The width of the interval is chosen to decrease as the logarithm of $\Delta H$ for $\xi_\| < L$ to minimize changes in $S_\mathrm{max}$ over the interval. A fixed width is used for $\Delta H L^{1/\nu_\|} < 2$, where $\xi_\| \gg L$.

\begin{figure}
	\includegraphics[width=0.45\textwidth]{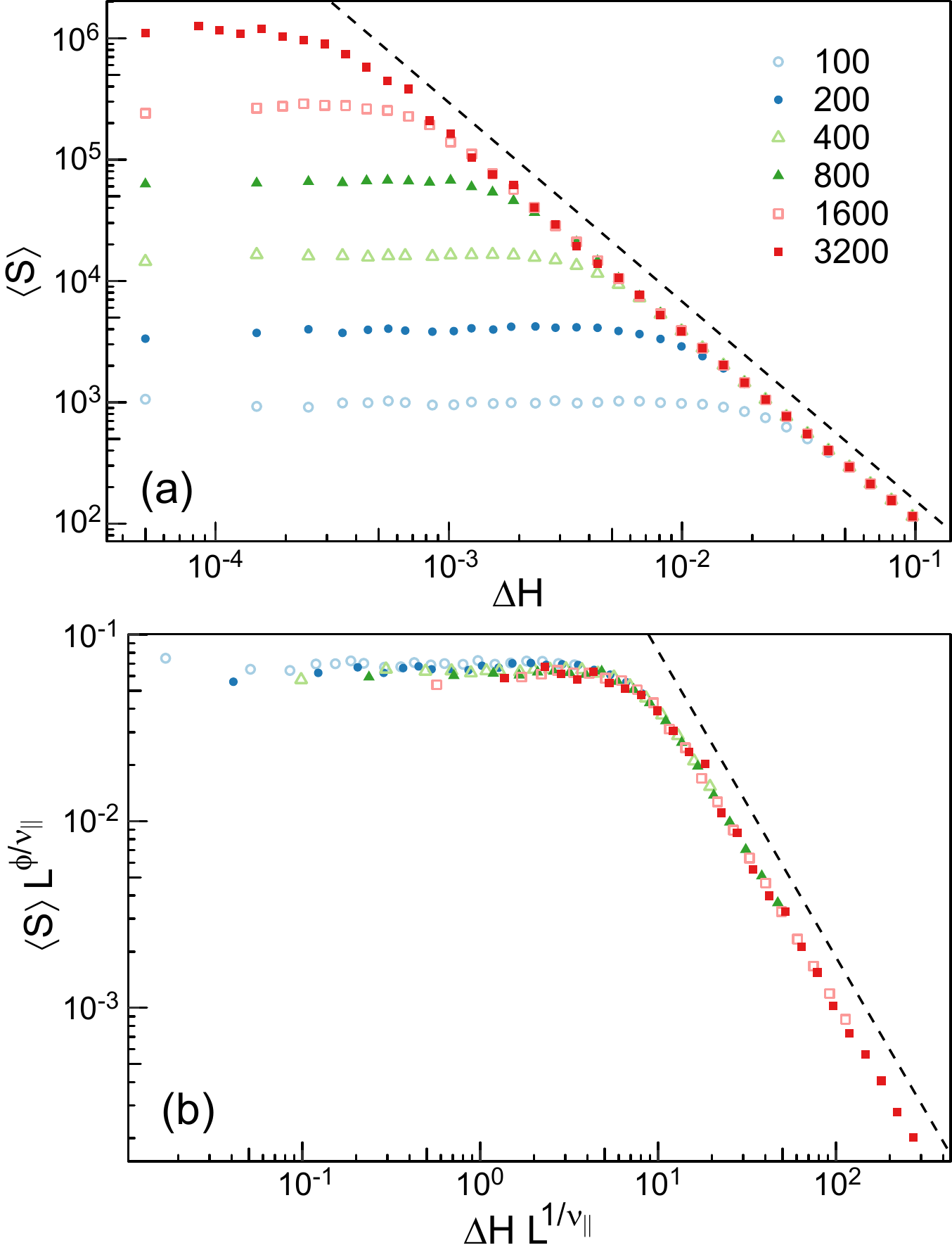}
	\caption{(a) The average volume of an avalanche $\langle S \rangle$  calculated at different values of $\Delta H$ for systems with size $L$ indicated in the legend. Note that the slope characterizing changes in $\langle S \rangle$ drops slightly for $ \Delta H > 10^{-2}$. (b) Scaling collapse for Eq. \eqref{eqn:ave_scaling} with values of $\phi = 1.64$ and $\nu_\| = 0.79$. Only data near the critical point, $\Delta H < 10^{-2}$, are included. Dashed lines in both panels indicate power-law scaling with $\phi=1.64$.}
	\label{fig:V1}
\end{figure}

Figure \ref{fig:V1}(a) shows the increase in $\langle S \rangle$ with decreasing $\Delta H$ at different $L$. For each $L$, $\langle S \rangle$ shows a power-law divergence, $\langle S \rangle \sim \Delta H^{-\phi}$, and then saturates at a value of $\Delta H$ that shrinks with increasing $L$. In the power-law regime where $L>\xi_\|$ , we can use Eq. \eqref{eqn:ave2} and $S_\mathrm{max} \sim \xi_\|^\alpha$ to determine a scaling relation:
\begin{equation}
\phi = \nu_\| \alpha (2-\tau) \ \ .
\label{eqn:phi}
\end{equation}
In the saturated region, $\xi_\| >L$, $S_\mathrm{max} \sim L^\alpha$ and $\langle S \rangle \sim L^{\phi/\nu_\|}$.

Given the above scaling behavior we can construct a finite-size scaling ansatz similar to Eq. \eqref{eqn:fs}:
\begin{equation}
\langle S \rangle \sim L^{\phi/\nu_\|} g\left(L^{1/\nu_\|} \Delta H \right) \ \ ,
\label{eqn:ave_scaling}
\end{equation}
where $g(x)$ is a new scaling function. In the asymptotic limit of $x \gg 1$, $g(x)$ will scale as $x^{-\phi}$ to reproduce the power law in Eq. \eqref{eqn:ave2}. Alternatively, when $x \ll 1$, $g(x)$ must approach a constant such that $\langle S \rangle \sim L^{\phi/\nu_\|}$.

Finite-size scaling holds only near the critical point. Close examination of Fig. \ref{fig:V1}(a) shows that the slopes of curves for all $L$ change for $\Delta H>10^{-2}$. This is consistent with later results in the main text and Appendix A that show critical behavior only for $\Delta H<10^{-2}$. Thus we include only fields within this range in finite-size scaling collapses. Note that $\langle S \rangle$ has saturated at $\Delta H > 10^{-2}$ for $L=100$. Results for $L=25$ and $50$ saturated even farther from the critical regime, and we do not include results for these small systems in this paper.

Figure \ref{fig:V1}(b) shows a scaling collapse of curves for different $L$ using $\nu_\| = 0.79$ and $\phi = 1.64$. Testing the sensitivity of the collapse to these parameters, we estimate uncertainties of $\nu_\| = 0.79 \pm 0.02$, consistent with Fig. \ref{fig:V1}, and $\phi = 1.64 \pm 0.04$. A direct measure of $\phi$ from Fig. \ref{fig:V1}(a) also yields $1.64 \pm 0.04$. Within error bars, this is consistent with the result from Ref. \cite{Ji1992}, $\phi = 1.71 \pm 0.11$.
 
A similar scaling procedure could also be performed on larger moments. However, the higher moments do not depend on any additional exponents, and they have increased sensitivity to the largest events, which are the hardest to sample.

\subsection{Avalanche distribution}

Having seen how the maximum avalanche volume $S_\mathrm{max}$ depends on $H$ and $L$, we next focus on the regime near $H_c$, where $\xi_\| > L$, and calculate the distribution of $S$ in order to isolate the exponents $\tau$ and $\alpha$. In this limit, $S_\mathrm{max}$ will no longer be limited by $\xi_\|$ but rather by $L$. We select avalanches that nucleated sufficiently close to the critical point such that $\xi_\| > L$ and designate the distribution as $P(S, L)$, dropping the dependence on field. Based on the length of the plateau in Fig. \ref{fig:V1}(b), we consider all nonspanning avalanches in the range $0 < \Delta H < 10 L^{-1/\nu_\|}$. This is consistent with the range where spanning avalanches dominate growth in Figs. \ref{fig:V0} and \ref{fig:V0.5}. Consistent scaling results were obtained for half and one tenth of this range.

\begin{figure}
	\includegraphics[width=0.45\textwidth]{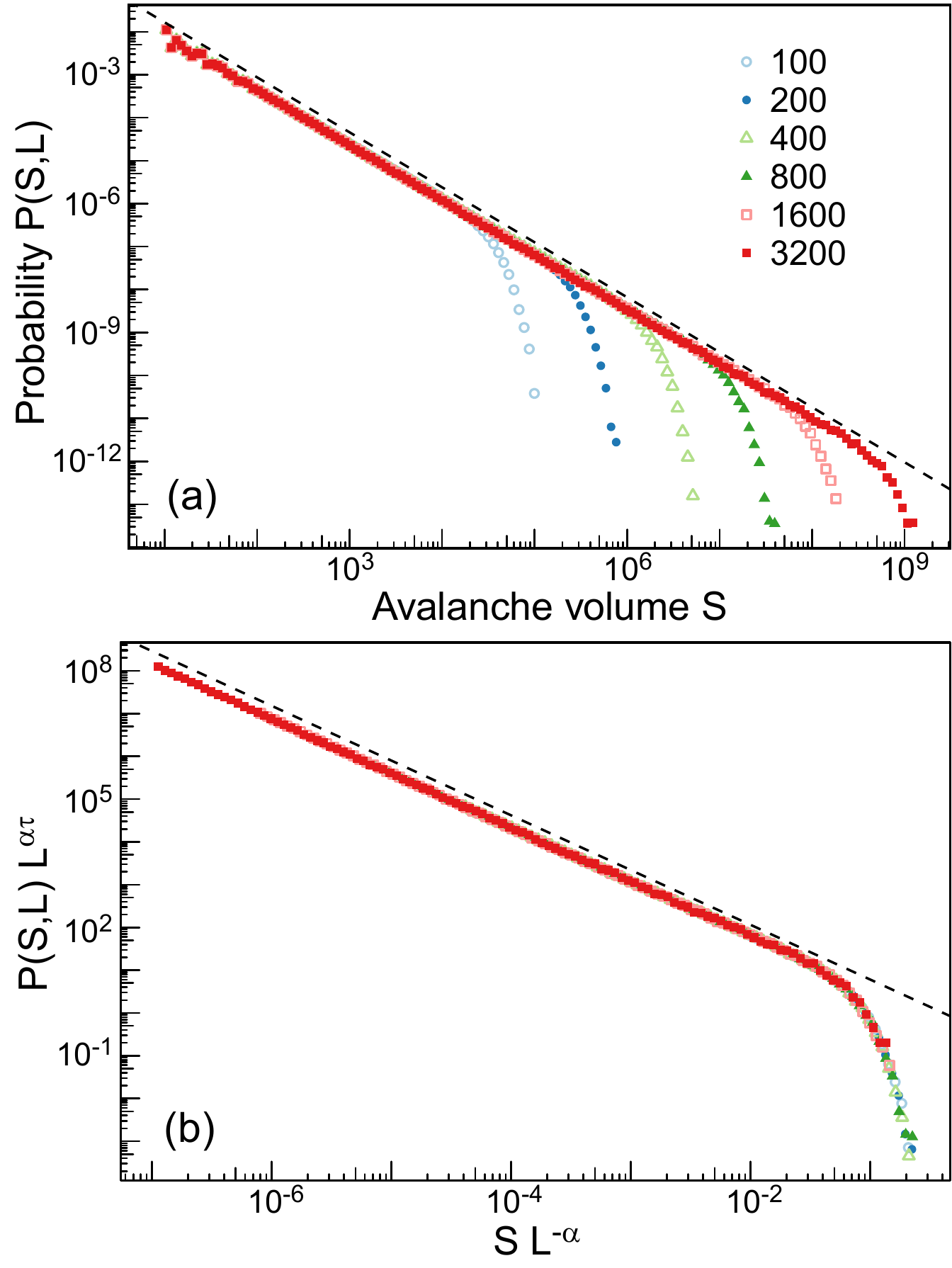}
	\caption{(a) The probability distribution of the volume of avalanches is calculated for the system sizes indicated in the legend at $\Delta H$ sufficiently close to the critical point such that $\xi_\| > L$, corresponding to $\Delta H < 10 L^{-1/\nu_\|}$. The dashed line represents a power law with $\tau = 1.28$. (b) The same data are collapsed by scaling with system size according to Eq. \eqref{eqn:ps2} with exponents $\tau = 1.28$ and $\alpha = 2.84$. Avalanches of $S < 10^3$ are excluded from the scaling. }
	\label{fig:V2}
\end{figure}

To calculate $P(S, L)$, avalanches are logarithmically binned by size, and the number of events in each bin is divided by the size of the bin before normalizing the distribution. The resulting distributions, seen in Fig. \ref{fig:V2}(a), have a clear power-law regime followed by a cutoff at a value of $S_\mathrm{max}$ that grows with increasing system size. As noted above, the fact that $P(S,L)$ is constant at low $L$ implies $\tau >1$. This is consistent with a direct evaluation of the slope, which gives $\tau = 1.28 \pm 0.01$. More accurate values are obtained by finite-size scaling.

The cutoff seen in Fig. \ref{fig:V1}(b) will depend only on the ratio of $S$ to $S_\mathrm{max}$, allowing us to write an expression for the distribution as
\begin{equation}
P(S, L) \sim L^{-\alpha \tau} f_p(S/L^{\alpha}) \ \ ,
\label{eqn:ps2}
\end{equation}
where $f_p(x)$ is another universal scaling function. For $x\gg 1$, $f_p$ goes to zero while for $x \ll 1$ one must have $f_p(x)\sim x^{-\tau}$ in order to recover the power-law scaling with $S$. This scaling should apply only for sufficiently large $S$ and $L$. In the previous section we found changes in behavior for $\Delta H < 10^{-2}$. Here we see evidence of deviations from scaling in avalanches with $S < 10^3$. The next section shows the discreteness of the lattice is important for these small avalanches, and thus they are excluded from finite-size scaling collapses. Including them does not significantly affect our best-fit estimates for exponents but affects the quality of the collapse.

Figure \ref{fig:V2}(b) shows a finite-size scaling collapse based on Eq. \eqref{eqn:ps2}. Based on the quality of the fit we estimate the values and uncertainties of the exponents as $\tau = 1.280 \pm 0.005$ and $\alpha = 2.84 \pm 0.02$. As noted above, this value of $\tau$ is between 1 and 2 and is consistent with direct evaluation of the slope in Fig. \ref{fig:V2}(a) and the value found for the RFIM in Ref. \cite{Ji1992}, $\tau = 1.28 \pm 0.05$. From Eq. \eqref{eqn:phi}, our values of $\tau$ and $\alpha$ predict $\phi = 1.62 \pm 0.05$, which is in agreement with the directly measured value.

The above results can be used to describe the amount of volume, $dV$, invaded over an interval of external field $d H$.
As shown in Appendix A, the rate of avalanche nucleation is proportional to the interfacial area $A$ and $A \propto L^2$ sufficiently close to $H_c$ and for sufficiently large $L$.
This is expected for interface motion since the interface moves into new
regions of space and the no-passing rule is obeyed \cite{Martys1991a,Narayan1993} and was verified for the case of fluid invasion \cite{Martys1991a}.
Note that very different behavior has been observed for critical behavior in sheared systems where the entire system is perturbed by internal avalanches and
they produce stresses that are not positive definite.
In these systems the rate of avalanches rises less rapidly than the system size
\cite{Salerno2012, Salerno2013, Lin2014a, Lin2015}.

If the rate of avalanches scales as $L^2$, then the total volume invaded over
an interval $dH$ scales as $dV \propto L^2 \langle S \rangle_A$.
Here, $\langle S\rangle_A$ indicates an average over all avalanches including spanning avalanches. Since the largest avalanches dominate $\langle S \rangle_A$ for $\tau<2$, spanning avalanches contribute most to $d V$. This explains why $F_s$ is near unity close to $H_c$ (Fig. \ref{fig:V0}).
If the system is at fields below the onset of finite-size effects or spanning avalanches, then $\langle S \rangle_A = \langle S \rangle$.
In this limit, the total volume invaded per unit area scales as
\begin{equation}
	\langle V \rangle /L^2 \sim \int \langle S \rangle dH \sim \int \Delta H^{-\phi} dH \sim \Delta H^{-\phi+1} \ \ .
	\label{eq:Vtot}
\end{equation}
As shown in Appendix A, this relation is valid sufficiently close to $H_c$ and for large $L$.

\subsection{Morphology of avalanches}

The above measurement of the exponent $\alpha$ allows us to estimate the anisotropy of correlations in the system. In $d$ dimensions, the largest avalanches will span an area $\xi_\|^{d-1}$ and reach a height $\xi_\perp$. From Eq. \eqref{eqn:xis} and the definition of $\chi$, this implies $\alpha = {d-1+\chi}$. Previous scaling relations assumed that $\chi = 1$ \cite{Ji1992, Martys1991} or the roughness exponent $\zeta$ \cite{Narayan1993}. Our numerical data imply $\chi = 0.84 \pm 0.03$ in three dimensions, which is midway between unity and previous measurements of $\zeta \sim 2/3$ \cite{Ji1992, Koiller2000}. This would imply that $\chi$ is a distinct exponent and there is a novel anisotropy in the RFIM not previously seen in other depinning systems. To test this, we next consider the morphology of individual avalanches.

In order to define the width $\ell_\|$ and height $\ell_\perp$ of an avalanche, we define a second moment tensor with components $l_{\alpha \beta}$, where $\alpha$ and $\beta$ represent the directions $x$, $y$, or $z$. Given an avalanche with a center of mass located at ($x_\mathrm{cm}$, $y_\mathrm{cm}$, $z_\mathrm{cm}$), we define the tensor components as:
\begin{equation}
l_{\alpha \beta} = \frac{1}{S} \sum_{i=1}^S \left( \alpha_i - \alpha_\mathrm{cm} \right) \left( \beta_i - \beta_\mathrm{cm} \right) \ \ ,
\end{equation}
where the summation over $i$ corresponds to a sum over all $S$ spins flipped by the avalanche. For avalanches that cross a periodic cell boundary, the positions of spins are unwrapped across the boundaries such that their position is measured relative to the original nucleation site.

Since periodic boundary conditions force the global motion to proceed in the $\hat{z}$ direction, avalanches will align with this orientation on average. However, an individual avalanche may nucleate and grow along a locally sloped region of the surface. In these instances, the avalanche's normal vector may not correspond to $\hat{z}$. To avoid biasing the results by assuming a local growth direction, we considered the eigenvalues of the second moment tensor, a method used in Ref. \cite{Nolle1996}. We associate $\ell_\perp^2$ with the smallest eigenvalue and $\ell_\|^2$ with the geometric average of the largest two eigenvalues \footnote{Using the arithmetic mean gives equivalent results.}. This decision is based on both the fact that $\xi_\perp < \xi_\|$ and the fact that growth is promoted along the local interfacial orientation due to the destabilizing effect of flipped neighbors. This definition will minimize the ratio $\ell_\perp /\ell_\|$ and therefore will also minimize estimates of $\chi$.

The corresponding eigenvectors of the second moment tensor indicate the direction of growth. At small scales, the orientation of the interface is arbitrary and the directions of the eigenvector $\hat{v}_\mathrm{min}$ associated with the smallest eigenvalue also varies. For self-affine surfaces the orientation is more sharply defined at large scales. Consistent with this, we find $\hat{v}_\mathrm{min}$ becomes more aligned with $\hat{z}$ as the size of the avalanche, $S$, increases relative to the size of the system. We quantify this alignment by the polar angle $\theta$ defined as $\cos{\theta} = \hat{v}_\mathrm{min} \cdot \hat{z}$. For $L = 3200$, avalanches with $S \sim 10^8$ have a root mean square (rms) deviation in angle from $\hat{z}$ of $\sim 6^\circ$. In contrast, the rms angular deviation grows to $\sim 39^\circ$ for small avalanches consisting of $10^3$ spins.

We also studied measures of $\ell_\|' \equiv \sqrt{l_{zz}}$ and $\ell_\perp' \equiv 
(l_{xx} l_{yy})^{1/4}$ that measure anisotropy relative to the periodic boundaries. The scaling behavior is similar but not as good. There appears to be a slight upwards shift in the height $\ell_\perp'$ of avalanches with increasing system size, particularly for smaller avalanches. As described in the following section, a larger system will ultimately reach a rougher final interface. 
This will increase the apparent $\ell_\perp'$ by mixing in $\ell_\|'$. We therefore focus on the principal component definition as it produced cleaner results.

\begin{figure}
	\includegraphics[width=0.45\textwidth]{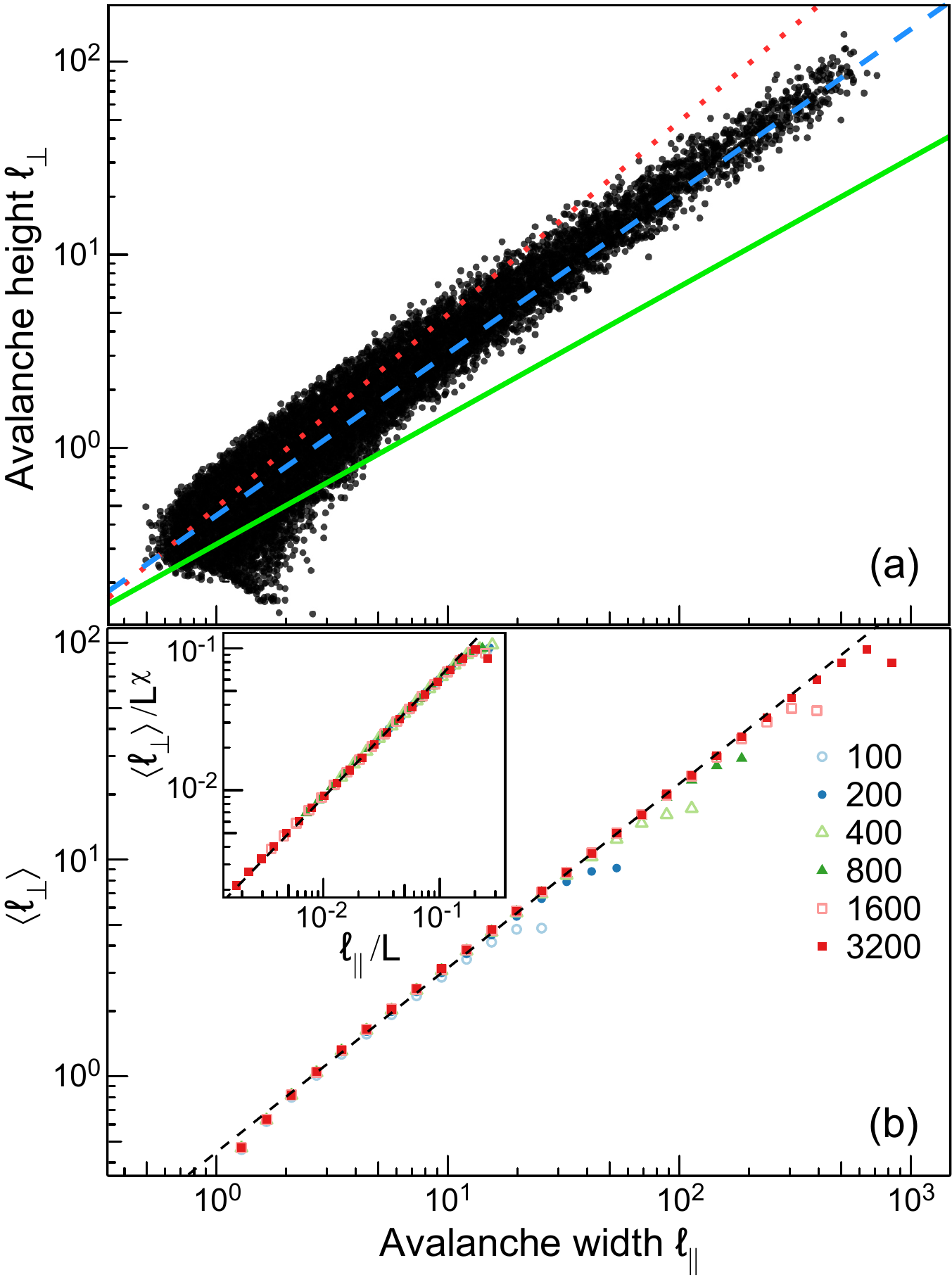}
	\caption{ (a) Sizes of individual avalanches along and perpendicular to the interface in systems of $L = 3200$ near $H_c$. Lines indicate power laws corresponding to $\chi = 1$ (dotted, red), 0.84 (dashed, blue), and 2/3 (solid, green). (b) Mean avalanche heights as a function of width for the system sizes indicated in the legend for $0 < \Delta H < 10 L^{-1/\nu_\|}$. Avalanches are binned by their width and the average height is calculated for the system sizes indicated in the legend. The dashed line has a slope of $\chi = 0.85$. In the inset, data for different $L$ with $\ell_\| > 5$ are collapsed by scaling the avalanche width by $L$ and the height by $L^\chi$ for $\chi = 0.85$.}
	\label{fig:M1}
\end{figure}

Values of $\ell_\|$ and $\ell_\perp$ were calculated for avalanches which nucleated sufficiently close to the critical point such that the largest avalanches were limited by system size rather than the correlation length. As in the previous subsection, the range was set to $0 < \Delta H < 10L^{-1/\nu_\|}$. In Fig. \ref{fig:M1}(a), $\ell_\perp$ is plotted as a function of $\ell_\|$ for a representative set of avalanches grown in a system of $L = 3200$. There is a broad spread among individual avalanches, but $\ell_\perp$ clearly grows sublinearly with $\ell_\|$, implying $\chi<1$ and thus that avalanches become proportionately flatter as they grow in size. The power-law rise is also clearly larger than previously measured values of the roughness exponent, $\zeta=0.67$, and consistent with our estimate of $\chi=0.84$ at the start of this section.

To accurately measure $\chi$, we binned avalanches by $\ell_\|$ and calculated the average value of $\ell_\perp$ for systems of a given $L$. Figure \ref{fig:M1}(b) shows that the mean height of an avalanche grows as a power of the width before being cut off due to finite-size effects. Note that the apparent power law changes for very small avalanches. The height can vary only in discrete steps of unity, and this will affect the scaling of avalanches with small $\ell_\perp$.  Based on the change in slope in Fig. \ref{fig:M1}(b) and the lack of scaling for $L < 100$, we include only avalanches with $\ell_\perp > 2$ in scaling collapses. This corresponds to $\ell_\| \gtrsim 5$ and $S \gtrsim 10^3$, which is consistent with the cutoff used in scaling $P(S,L)$.

In the critical region, results for $\ell_\|$ and $\ell_\perp$ should collapse when each is scaled by an appropriate power of the system size $L$. The maximum width of an avalanche is limited by $L$ due to the finite box size and the restriction that an avalanche is nonspanning. The corresponding maximum height an avalanche can attain must scale as $L^\chi$. The inset in Fig. \ref{fig:M1}(b) shows that curves for different $L$ collapse when each length is scaled by its maximum value. Varying $\chi$, we find a collapse is achieved for the range of $\chi = 0.85 \pm 0.02$. Alternatively, one could bin by $\ell_\perp$ and average $\ell_\|$. This process produces similar values of $\chi$.

\begin{figure*}
	\includegraphics[width=\textwidth]{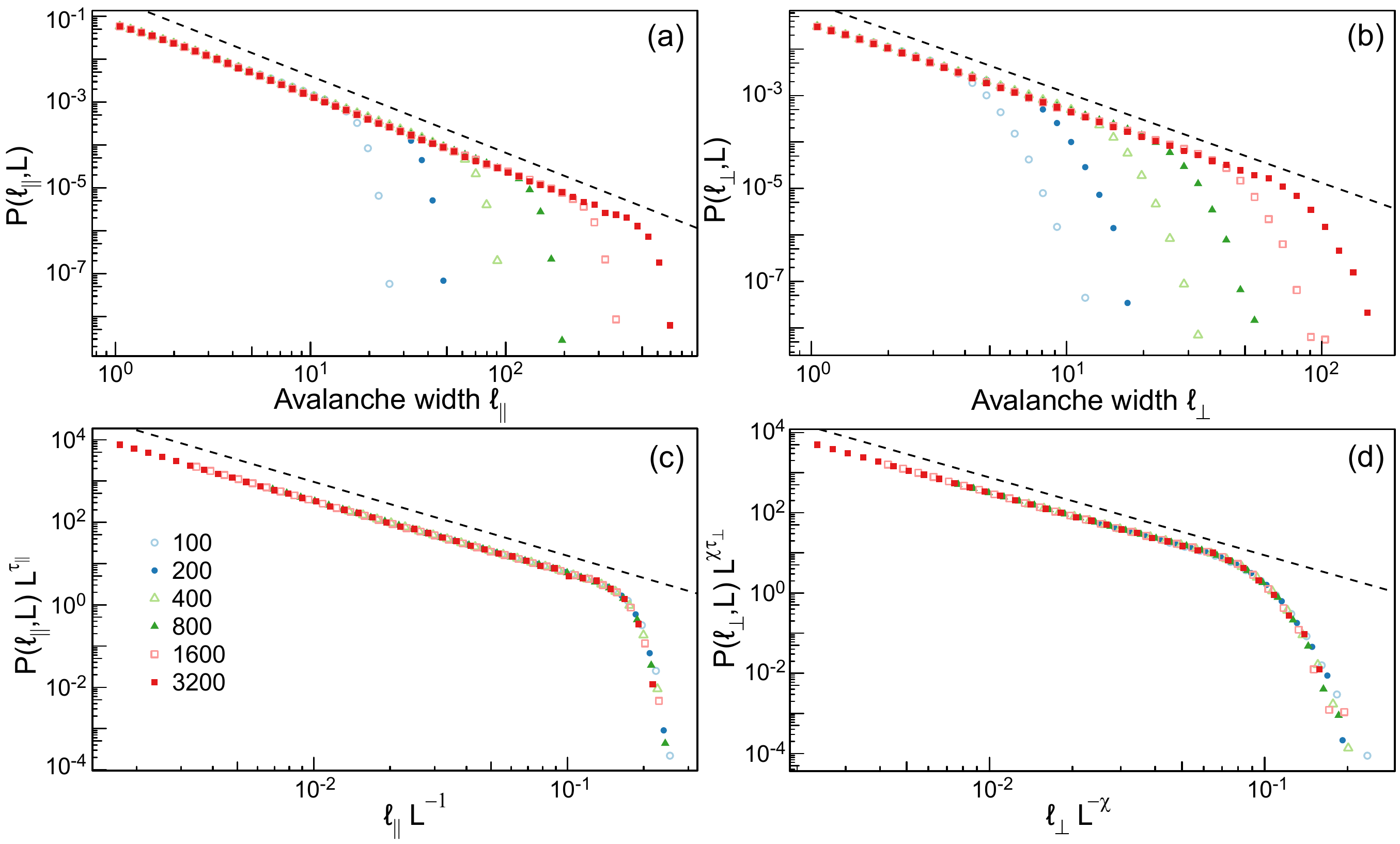}
	\caption{The probability distributions of the (a) width and (b) height of avalanches for values of $L$ indicated in the legend of panel (c) that grew at fields near the critical regime $0< \Delta H <10 L^{-1/\nu_\|}$. The distributions in panels (a) and (b) are collapsed by scaling with system size using exponents $\chi = 0.85$, $\tau_\| = 1.79$, and $\tau_\perp = 1.94$ in panels (c) and (d), respectively. Data for $\ell_| < 5$ or $\ell_\perp < 2$ are excluded from scaling. Dashed lines in each panel indicate the power law determined from finite-size scaling. }
	\label{fig:M2}
\end{figure*}

As for the distribution of avalanche volumes, one can also define the probability for a given linear dimension at a given $H$ and $L$, $P(\ell, H, L)$ where $\ell$ is either $\ell_\|$ or $\ell_\perp$. These distributions are expected to decay as a power law with an exponent $\tau_\|$ or $\tau_\perp$. This power law will persist only up to a maximum cutoff set by either the correlation length or the system size. As in Fig. \ref{fig:V2}(a), we focus on the critical distribution $P(\ell,L)$ at $H$ close enough to the critical point that avalanches are limited by the finite system size rather than the correlation length. Figures \ref{fig:M2}(a) and \ref{fig:M2}(b) show $P(\ell_\|, L)$ and $P(\ell_\perp, L)$, respectively. The distributions are seen to decay with different exponents before being cut off at a threshold that grows with $L$.

Following Eq. \eqref{eqn:ps2}, one can construct finite-size scaling equations for the distributions of the heights and widths of avalanches. As demonstrated in Fig. \ref{fig:M1}(b), the maximum width of an avalanche will scale in proportion to $L$, and the maximum height will scale in proportion to $L^\chi$. Thus $\alpha$ in Eq. \eqref{eqn:ps2} is replaced by 1 or $\chi$ for $\ell_\|$ and $\ell_\perp$, respectively. Figures \ref{fig:M2}(c) and \ref{fig:M2}(d) show finite-size scaling collapses for both quantities. By varying the choice of exponents we determined the data are consistent with $\tau_\| = 1.79 \pm 0.01$, $\tau_\perp = 1.94 \pm 0.02$, and $\chi = 0.85 \pm 0.01$. In both collapses, we exclude avalanches with length scales $\ell_\| < 5$ or $\ell_\perp < 2$. At smaller length scales, the measurements are affected by the discreteness of the lattice and they do not follow the power-law scaling in Fig. \ref{fig:M1}(a).

The $\tau$ exponents are not independent and can be related to each other as derived in Ref. \cite{Nolle1996}. In Fig. \ref{fig:M1}(a), one can see that, on average, individual avalanches exhibit the same anisotropy as the correlation lengths, typically $\ell_\perp \sim \ell_\|^\chi$. Thus we assume this scaling will hold when considering the statistics of many avalanches. For length scales $\ell_\| < \xi_\|$ and $\ell_\perp < \xi_\perp$ we can equate the probability that avalanches are in a range with corresponding values of $\ell_\|$ and $\ell_\perp$: $P(\ell_\|) d\ell_\| \sim P(\ell_\perp) d\ell_\perp$. Using this expression, one can derive a scaling relation relating the exponents $\tau_\|$ and $\tau_\perp$ to $\chi$:
\begin{equation}
\chi = \frac{\nu_\perp}{\nu_\|} = \frac{1 - \tau_\|}{1 - \tau_\perp} \ \ .
\label{eqn:txz}
\end{equation}
Using our estimates of $\tau_\|$ and $\tau_\perp$, this yields an estimate of $\chi = 0.84 \pm 0.02$, again consistent with our findings. Similarly, one can relate the rate of avalanches over a small interval of volumes, $dS$, to the rate of avalanches over a small interval of widths, $d\ell_\|$, and derive a relation between $\tau$ for the volume distribution and $\tau_\|$ \cite{Nolle1996}:
\begin{equation}
\tau = \frac{d-2+\chi +\tau_\|}{d - 1 + \chi} \ \ .
\label{eqn:ts}
\end{equation}
Plugging in our values for $\tau_\|$ and $\chi$ we find a prediction of $\tau = 1.28 \pm 0.01$ in strong agreement with the value directly measured in Fig. \ref{fig:V2}.

One other scaling relation is implied by our results. As noted in the previous section and Appendix A, the ratio $\langle V \rangle /L^2 \propto \Delta H ^{-(\phi-1)}$ near $H_c$.
This is proportional to the average height of the external interface because the volume left behind in bubbles is a small constant fraction of the total volume.
The average height of the interface should be at least as big as the height of the largest avalanches. Since $\ell_\perp \propto \Delta H ^{-\nu_\perp}$, 
this implies $\phi-1 \geq \nu_\perp$.
Within our error bars, our directly measured values of $\phi-1 = 0.64 \pm 0.04$ and $\nu_\perp = 0.67 \pm 0.02$ are consistent with this relation and suggest
that
\begin{equation}
	\phi = 1+ \nu_\perp \ \ .
	\label{eq:scale2}
\end{equation}
The numerical results in Refs. \cite{Ji1992,Martys1991} were consistent
with $\chi=1$, and they tested a scaling relation $\phi=1+\nu$ that is equivalent to Eq. \eqref{eq:scale2} in that limit.

Overall, in these past two sections we proposed and tested a theory of avalanches that accounts for the anisotropy in correlation lengths. From these results, we identified several measures of $\chi$ confirming it is distinct from both 1 and the previously measured roughness exponent. Next we explore how this scaling changes for spanning avalanches.

\subsection{Spanning avalanches}

Defining the morphology of a spanning avalanche is complicated. Having percolated, each flipped spin has different paths connecting it to a nucleation site in any periodic image. There is no longer a well defined reference point to define the lateral position $(x,y)$ of a flipped spin. Therefore, neither the second moment tensor $l_\mathrm{\alpha \beta}$ nor its eigenvalues are uniquely defined. However, the height of an avalanche can still be estimated using the metric $\ell_\perp' = \sqrt{l_{zz}}$ because the calculation of $l_{zz}$ is not affected by the periodicity of the lateral boundary conditions. As discussed above, this is not an ideal measure of the height for small avalanches. However, spanning avalanches are large and sample the global slope of the interface. Therefore spanning avalanches are expected to closely align with $\hat{z}$ such that $\ell_\perp'$ is a reasonable measure of their height.

From the definition of $\chi$ and Eq. \eqref{eqn:xis}, the height of the typical nonspanning avalanche is expected to grow as a power of $S$ with exponent $\chi/\alpha \approx 0.3$. Since spanning avalanches detect the finite boundaries there is no guarantee that they will obey the same scaling.

\begin{figure}
	\includegraphics[width=0.45\textwidth]{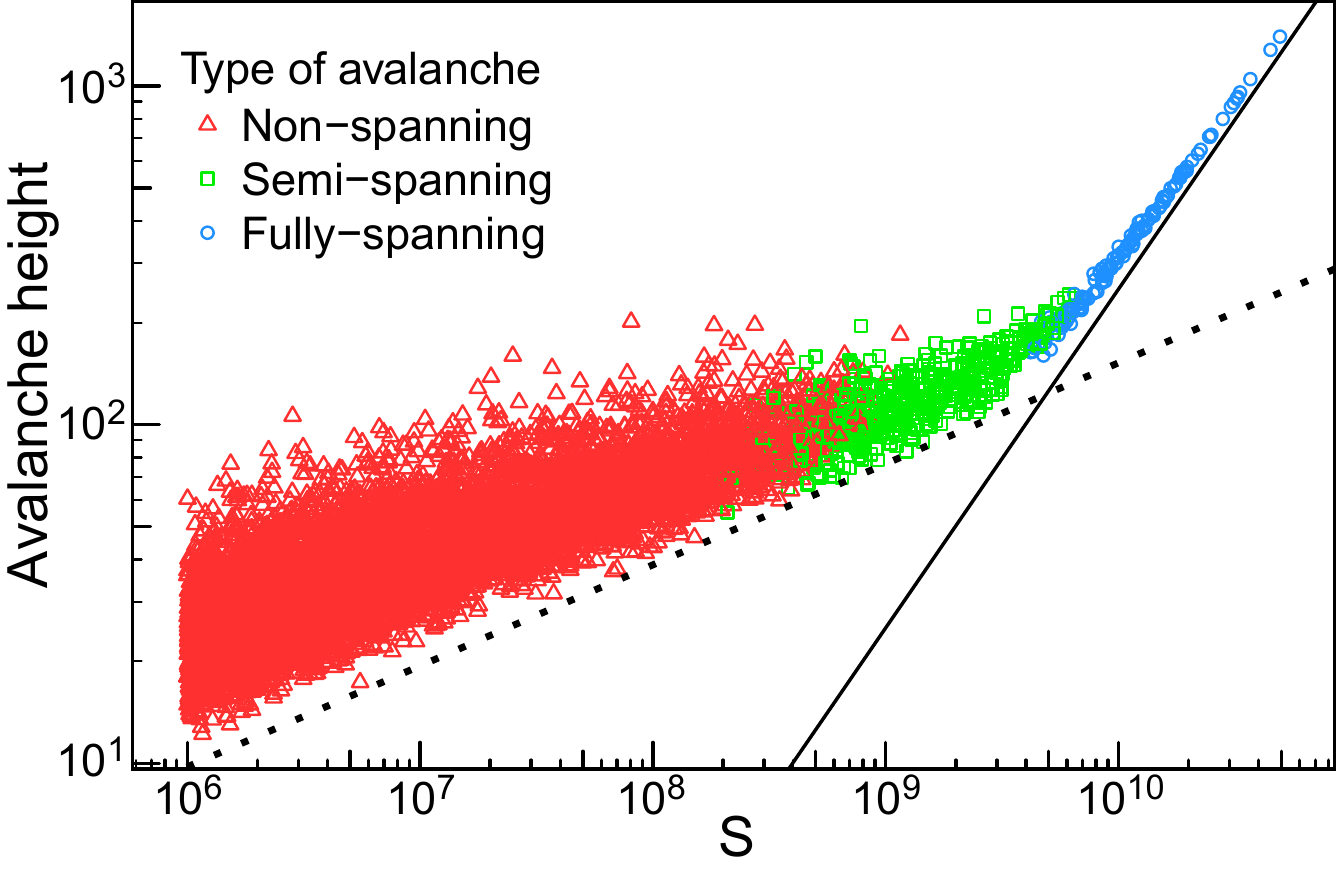}
	\caption{ Height of an avalanche in the $\hat{z}$ direction, $\ell_\perp'$, as a function of volume. Nonspanning avalanches are red triangles, semispanning avalanches are green squares, and fully spanning avalanches are blue circles. Straight lines have a slope of $\chi/\alpha \approx 0.3$ (dotted) and 1.0 (solid).}
	\label{fig:M3}
\end{figure}

To test for deviations from scaling, we calculate $\ell_\perp'$ and $S$ for all avalanches nucleated close to the critical point for $L = 3200$. As above, we considered fields in the range $0 < \Delta H < 10L^{-1/\nu_\|}$ such that $\xi_\| > L$. In Fig. \ref{fig:M3}, $\ell_\perp'$ is plotted as a function of $S$ for a sample of avalanches of size $S > 10^6$. Data are colored by the degree of spanning for each avalanche. Although there is a large amount of scatter for $S < 5 \times 10^9$, $\ell_\perp'$ is seen to grow as a power of $S$. These data are consistent with the predicted exponent of $\chi/\alpha$.  Above this scale, the height starts to grow in proportion to $S$. This threshold approximately corresponds to the division between semi- and fully spanning avalanches.

The distinct scaling of fully spanning avalanches seen in Fig. \ref{fig:M3} can be understood in the context of their definition. Once an avalanche grows to have a footprint of $L^2$, the width of the avalanche is fixed at the box size. Growth in the total volume must then be proportional to an increase in height. As seen in Fig. \ref{fig:Method0}, a fully spanning avalanche can have a much larger aspect ratio of height to width than a semispanning avalanche. One might also anticipate this change in scaling due to similarities of fully spanning avalanches to depinning. Once an avalanche grows to have a footprint $L^2$, flipped spins cover the entire cross section. Although it is still possible for some parts of the  multivalued interface to be pinned, they are usually left behind in bubbles and the external interface is totally renewed. Thus, fully spanning avalanches are more representative of motion above the depinning transition, and their scaling is not relevant to the behavior at $H < H_c$ of interest in this paper.

This interpretation is verified in Appendix B where we analyze the finite-size scaling of the probability distribution of $S$ including semispanning as well as all avalanches.
In a system of linear size $L$, we find the largest semispanning events scale as $L^\alpha$ implying that their height scales as $L^\chi$ with the same exponents found for nonspanning events.
In contrast, the largest fully spanning events scale as $L^3$ implying their height scales as $L$, which is consistent with Fig. \ref{fig:M3}.
Note that this scaling is affected by our growth protocol.
Avalanches are censored if their growth is interrupted by hitting the top of the box of height $2L$. This produces an artificial maximum height that scales with $L$. Thus the observation that the height of fully spanning avalanches scales as $L$ is consistent with them representing depinning events that are related to growth above $H_c$ and they could sweep out much larger volumes if growth was allowed to continue.

\subsection{Total interface roughness}

Having identified a distinct anisotropy in avalanches, we now explore how this impacts the morphology of the advancing interface.
One can study the statistical properties of stable interfaces without resolving all preceding avalanches.
Therefore, we were able to use our alternative growth protocol where we simply flip spins until a stable interface is reached at a fixed field.
The no-passing rule guarantees that this stable interface is independent of the growth rules, and efficient parallelization of the code allows us to study larger system sizes, up to $L = 12800$. In the following we identify the interface position with the set of flipped spins on the external interface that are adjacent to unflipped spins. Using the unflipped spins gives nearly identical results, particularly at large scales.

\begin{figure}
	\includegraphics[width=0.45\textwidth]{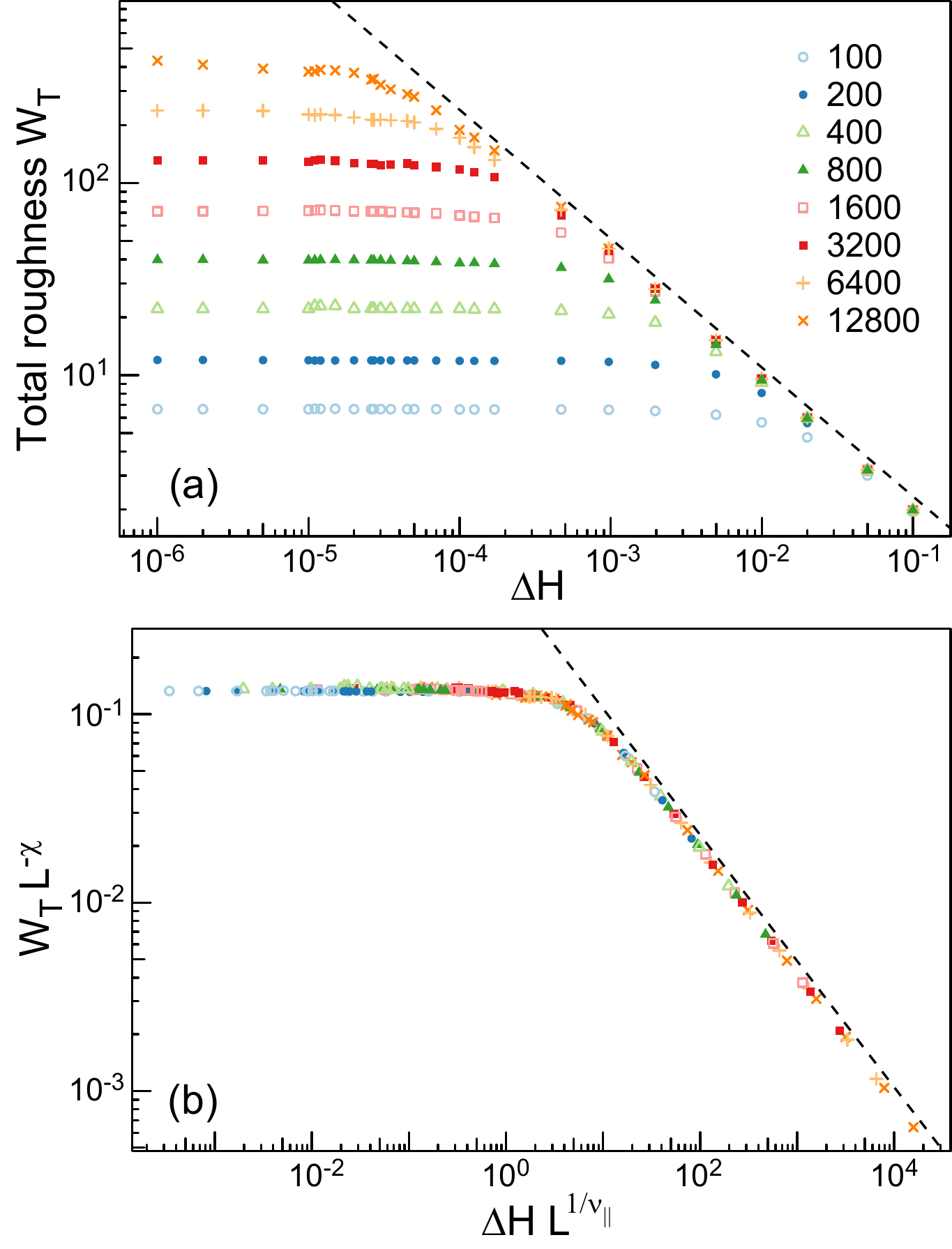}
	\caption{(a) The rms height variation of the external interface $W_T$ as a function of external field is shown for the values of $L$ indicated in the legend. A dashed line of slope $\nu_\perp = 0.67$ is shown. (b) The same data are shown after scaling the axes with powers of system size using $\chi = 0.85$ and $\nu_\| = 0.79$. }
	\label{fig:SP2}
\end{figure}

We first explore the total interface roughness, $W_T(L, \Delta H)$, defined as the root mean squared (rms) variation in the height $h(x,y)$ of all interfacial spins on the external interface. Note that the height is multivalued, and all spins at a given $x$ and $y$ are included in calculating $W_T$. 
Figure \ref{fig:SP2}(a) shows how $W_T$ grows as $H$ approaches $H_c$ for different $L$. The interface starts as a flat plane with $W_T = 0$ at large $\Delta H$. As $\Delta H$ decreases, the interface advances and roughens.
For each $L$,
$W_T$ grows as an inverse power of $\Delta H$ and then saturates. Saturation occurs at a larger roughness and smaller $\Delta H$ as $L$ increases.

$W_T$ is expected to grow at least as rapidly with decreasing $\Delta H$ as the height of the largest avalanches, i.e., $\xi_\perp$. Smaller or larger variations could be observed if successive events were anticorrelated or correlated on scales of order $\xi_\|$ to spread or concentrate growth. Assuming there are no such correlations, we predict $W_T \sim \Delta H^{-\nu_\perp}$ from Eq. \eqref{eqn:xis}. Fitting the power-law region in Fig. \ref{fig:SP2}(a) gives $\nu_\perp = 0.67 \pm 0.02$. Given our measured value of $\nu_\|$ this implies $\chi = 0.85 \pm 0.04$ in close agreement with our other results for $\chi$.

The finite-size saturation of $W_T$ in Fig. \ref{fig:SP2}(a) can be understood in terms of the scaling of the maximum height of an avalanche with $L$. The maximum height of a nonspanning avalanche is seen in Fig. \ref{fig:M2} to grow in proportion to $L^\chi$. This would suggest $W_T$ will saturate at a value proportional to $L^\chi$. Close to $H_c$, fully spanning avalanches will also contribute to the structure of the interface. As discussed above, fully spanning avalanches have a height that scales as $L$. However, as seen in Fig. \ref{fig:Method0}, the height of a fully spanning avalanche is not necessarily correlated with the interface width. If the entire interface is advanced a fixed distance, it does not change the width of the interface. Only the external topology of a fully spanning avalanche is relevant to $W_T$. Assuming fully spanning avalanches do not alter the scaling with $L$, we propose the following scaling ansatz for $W_T$:
\begin{equation}
W_T \sim \xi_\perp f_W\left( L/\xi_\| \right) \ \ ,
\label{eqn:w2}
\end{equation}
where $f_W(x)$ is a new scaling function. To satisfy the limiting scaling behavior, $f_W(x)$ goes to a constant for $x\ll 1$ and scales as $x^\chi$ for $x\gg 1$.

Figure \ref{fig:SP2}(b) shows a finite-size scaling collapse of the data in Fig. \ref{fig:SP2}(a). As before, we restrict data to $\Delta H < 10^{-2}$ because the lower fields do not represent critical behavior. Good scaling collapses are obtained for $\chi =0.85 \pm 0.01$ and  $\nu_\| = 0.79 \pm 0.02$. These values are consistent with those found above. It is worth noting that $W_T$ saturates at $\Delta H \approx 10 L^{1/\nu_\|}$ for different $L$. This onset of finite-size saturation in $W_T$ occurs at about the same field as the onset of finite-size effects in $\langle S \rangle$ shown in Fig. \ref{fig:V1}(b). This is evidence that fully spanning avalanches do not alter the scaling of the total interfacial width as assumed by the ansatz in Eq. \eqref{eqn:w2}.

\subsection{Test of self-affine scaling}

The anisotropy in avalanches and the fact that $W_T$ grows sublinearly with $L$ are consistent with self-affine scaling.
For a self-affine surface, the rms variation in height $W$ over an $\ell$ by $\ell$ square in the $x$-$y$ plane
scales as:
\begin{equation}
	W(\ell, H,L) \sim \ell^\zeta \ \ ,
\label{eqn:w0}
\end{equation}
where $\zeta$ is the roughness or Hurst exponent \cite{Meakin1998}.
For a finite system, one expects the total roughness to scale as $L^\zeta$, implying $\zeta=\chi$ from the results above.
This is inconsistent with past values of $\zeta$ and we
now test this scaling.

One complication is that the surface height $h(x,y)$ is not a single-valued function, as usually assumed for self-affine surfaces.
In order to circumvent regions of strong pinning, the system is capable of lateral growth that produces overhangs in the external interface.
Previous studies have shown that these overhangs have a characteristic size that diverges as $\Delta \rightarrow \Delta_c$ \cite{Koiller2000}.
We focus on $\Delta=1.7$ to reduce their size, but found similar behavior for $\Delta=2.1$, 2.0, 1.5, and 1.0. 

To calculate $W$, the periodic $x$-$y$ plane was divided into square cells of edge $\ell$.
For each cell, all interfacial sites contained in the projected area were used to calculate the rms variation in height over the cell.
Taking an average over $N_\ell$ cells of size $\ell \times \ell$ gives the scale dependent roughness:
\begin{equation}
W(\ell,\Delta H,L) = \frac{1}{N_\mathrm{\ell}} \sum_{i} \sqrt{\langle \left(z - \langle z \rangle\right)^2 \rangle} \ \ ,
\end{equation}
where the summation is across all $N_\mathrm{\ell}$ cells and the angular brackets represent averages within each cell \cite{barabasi1995fractal}.

\begin{figure}
	\includegraphics[width=0.45\textwidth]{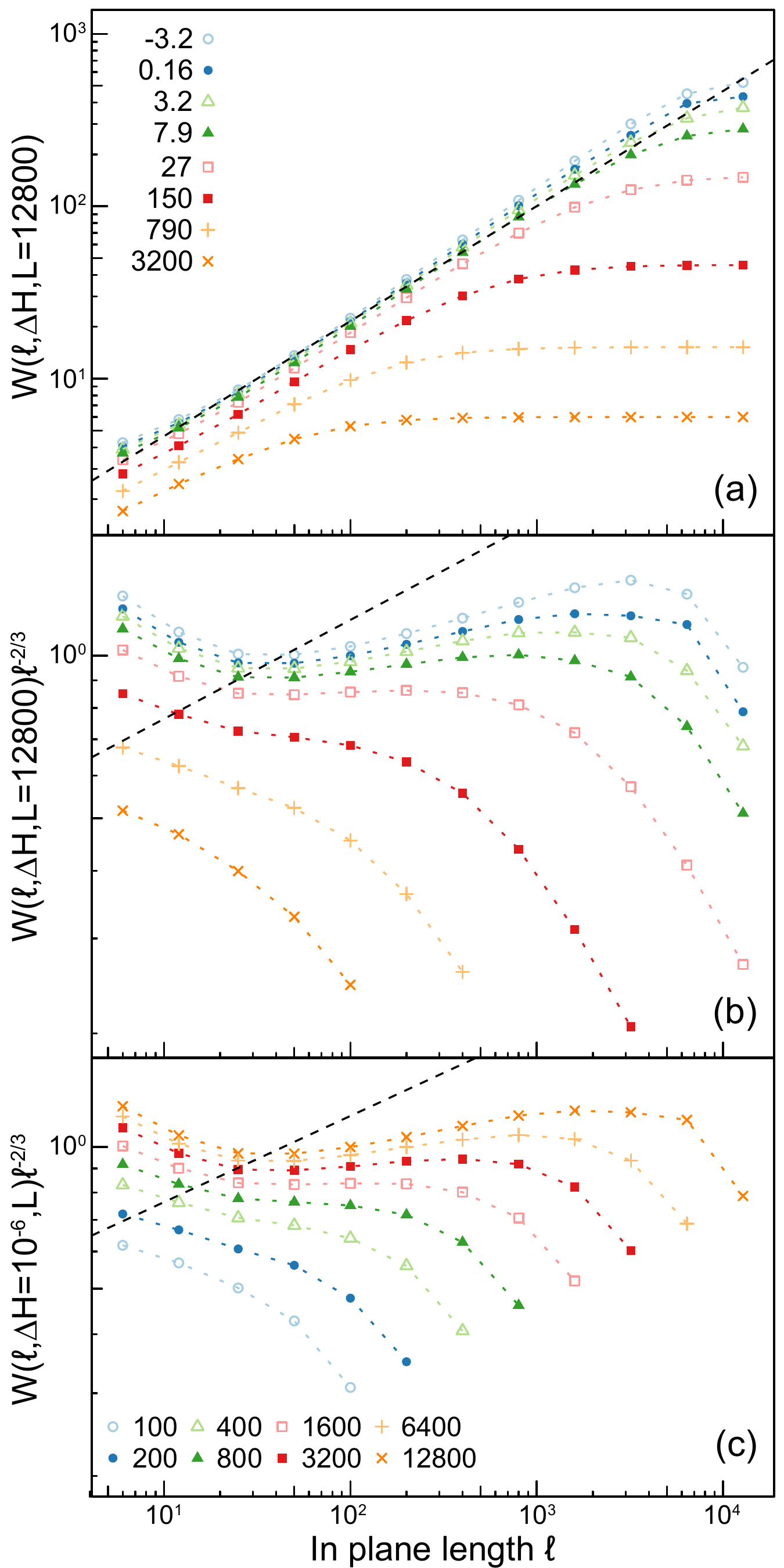}
	\caption{ (a) The rms fluctuation in height $W(\ell, \Delta H,L)$ is calculated for a system of size $L = 12800$. Values of $\Delta H L^{1/\nu_\|}$, rounded to two significant digits, are indicated in the legend.
	A dashed line is drawn with a slope of $2/3$. (b) The same values of $W$ are divided by the mean-field power law $\ell^{2/3}$. A line is included that corresponds to $\zeta = 0.85$.
(c) The variation of $W$ with $\ell$ at $\Delta H=10^{-6}$ for the indicated values of $L$. Once again $W$ is divided by the mean-field behavior, and the line corresponds to $\zeta=0.85$.} \label{fig:w0}
\end{figure}

Figure \ref{fig:w0}(a) shows how $W(\ell, \Delta H,L)$ evolves during growth for $L=12800$. The curves rise more slowly with $\ell$ below a lower scale $\ell_\mathrm{min}$. This is associated with the size of the overhangs mentioned above, which lead to a finite width even for $\ell=1$. For both single and multivalued interfaces we find different scaling with $\ell$ below $\ell_\mathrm{min} \sim 25$.
For larger $\ell$, $W$ appears to rise as a power-law before saturating at a roughness that grows as $H$ approaches $H_c$.
This asymptotic value corresponds to $W_T(L, \Delta H)$ (Fig. \ref{fig:SP2}).

Closer inspection shows that the power law rise in $W$ with $\ell$ is of limited range and has a power-law exponent that depends on $\ell$ and $\Delta H$. To reveal this, $W$ is multiplied by $\ell^{2/3}$ and replotted in Fig. \ref{fig:w0}(b).
This would produce horizontal lines if $\zeta$ had the mean-field value of $2/3$
\cite{Grinstein1983}.
For small $\Delta H$ ($L^{1/\nu} \Delta H < 10$), there may be a factor of 30 over which the curves are straight and thus follow a power law.
However, there is a steady rise in the slope with $\Delta H$.
Figure \ref{fig:w0}(c) shows similar scaled plots of $W$ at the critical field for different $L$. Once again there is a power-law region that grows with $L$, but
no clear saturation in slope that would indicate an approach to the limiting $\zeta$.
For $\Delta H= 10^{-6}$ and $L=12800$, the slope has risen to about 0.75, which is substantially above the mean-field exponent but well below $\chi$ (straight dashed line).

The results in Fig. \ref{fig:w0} imply either that growing interfaces are still affected by finite system size or that the interfaces are not simply self-affine. Some growth processes produce multiaffine surfaces where different moments of the height variation produce different scaling exponents \cite{Meakin1998}. To test this, we studied the scaling of the mean absolute value of height changes and the fourth root of the fourth power of height variations. The same scaling behavior was observed as for the rms height change.
We also examined the scaling of single-valued interfaces corresponding to the highest spin at a given $x,y$ or the average spin height at each $x,y$.
Similar to past results \cite{Nolle1993, Nolle1994,Koiller2000}, we see the roughness differs slightly at small $\ell$. However, the single-valued interfaces show the same shift in power law with $\Delta H$ and $L$, with similar exponents.

Another possibility is that depinning avalanches erase memory of the initial interface orientation and that subsequent growth is self-affine relative to the new local orientation.
To test this we used a technique like that used in finding the normal component of avalanches.
For each interface section of size $\ell \times \ell$ normal to the global growth direction, the moment tensor was calculated, and the smallest eigenvalue was taken as the height variation.
This approach maximizes the apparent $\zeta$ because it reduces the roughness at small $\ell$ and has little effect at large $\ell$.
We found that the range of power-law scaling was smaller using this metric and that the exponent showed a similar increase with decreasing $\Delta H$ and increasing $L$. The largest value of the apparent slope increased only to $0.79$, which is still smaller than $\chi$.

The origin of the change in apparent exponent seems to be the variation in roughness at small $\ell$ with increasing $L$ and decreasing $\Delta H$.
Growing interfaces often follow the Family-Vicsek relation \cite{Family1985}.
At each position, the roughness grows as $\ell^\zeta$, and then saturates.
The value of $\ell$ where saturation occurs grows as the interface advances,
as does the total roughness.
Figure \ref{fig:w0} shows similar behavior with decreasing $\Delta H$ with one important difference. The value of $W$ at points before saturation rises steadily as the interface advances, while Family-Vicsek scaling assumes that the small $\ell$ roughness is unchanged.

\begin{figure}
	\includegraphics[width=0.45\textwidth]{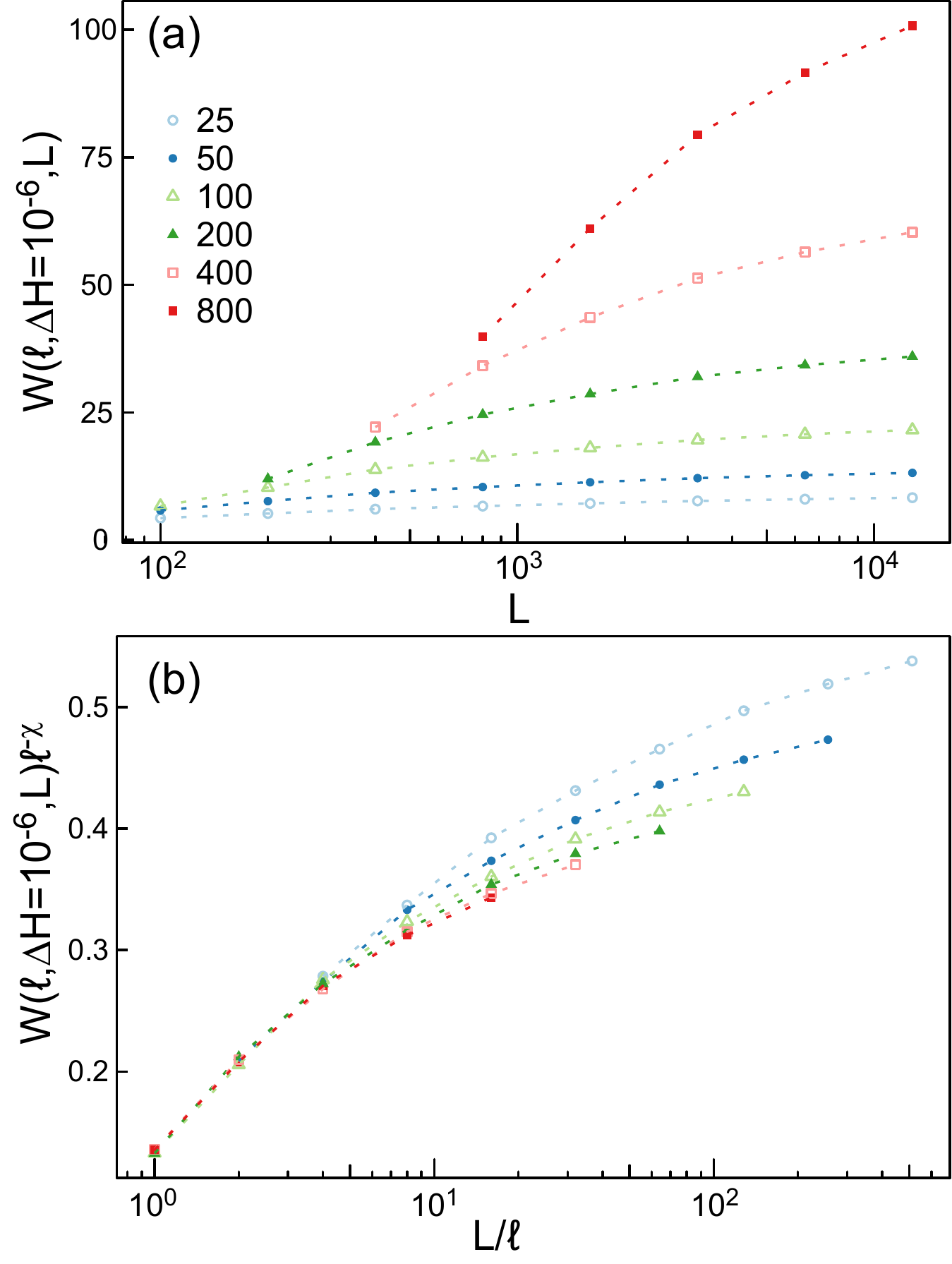}
	\caption{(a) The roughness at $\ell$ equal to the values indicated in the legend is calculated as a function of $L$ at a field of $\Delta H = 10^{-6}$. (b) The values of $W$ and $L$ in panel (a) are normalized by $\ell^\chi$ and $\ell$, respectively.}
	\label{fig:Wsmall}
\end{figure}

Figure \ref{fig:Wsmall} shows how the roughness at a fixed $\ell$ varies with $L$ close to the critical point ($\Delta H=10^{-6}$).
The previous section showed that $W_T=W(\ell = L,\Delta H=0,L) \propto L^\chi$.
If $W(\ell,\Delta H=0,L) \propto \ell^\chi$ with no dependence on $L$,
then one would have $W(L,0,L)/W(\ell,0,L) \propto (L/\ell)^\chi$, and the plots in Fig. \ref{fig:w0} would be power laws with the same slope.
However, $W(\ell,0,L)$ grows with $L$, and this decreases the ratio
$W(L,0,L)/W(\ell,0,L)$ and thus the apparent exponent.
If $W$ rose as a power of $L$, there would be a persistent difference between
$\zeta$ and $\chi$.
However, the linear-log plot in Fig. \ref{fig:Wsmall} shows that the growth
in $W$ is slower than logarithmic.
This supports the conclusion that $\zeta$ converges to $\chi$ in the
	thermodynamic limit, and the variation with $L$ in $W$ at small $\ell$
is large enough to explain the apparent difference of $\sim 0.1$ between $\zeta$ and $\chi$ for our system sizes.

It is interesting to compare our results to previous studies.
Past simulations for the RFIM \cite{Koiller2000,Ji1992} were consistent with
$\zeta=0.67\pm 0.3$, but used $L \leq 768$ and saw only
scaling to about $\ell=300$.
Our results for comparable $L$ give similar apparent slopes,
but data for larger $L$ reveal that this slope is not the limiting value.
Studies of models with explicitly broken symmetry and single-valued interfaces have found $\zeta=0.753 \pm 0.002$ using systems with $L \leq 400$ \cite{Rosso2003}.
It is possible that breaking symmetry leads to a reduction in $\zeta$, but it would be interesting to verify this with larger simulations.
Indeed, epsilon expansion calculations for single-valued models yielded
$\zeta=0.67$ and $0.86$ at first and second order, and estimated a converged
value of $0.82 \pm 0.1$ \cite{Chauve2001}.
It is interesting that the last prediction is close to the value of $\chi$ found here.

\subsection{Overhangs}

In the above section, we found that the interface continues to roughen on length scales $\ell < \xi_\|$ as $H$ increases, complicating measurement of the roughness exponent. This section quantifies the contribution of overhangs to the roughness as systems approach the critical point and shows that their contribution to the surface roughness becomes irrelevant as $L\rightarrow \infty$.

To identify multivalued locations on the interface, we first find the minimum and maximum height of the interface at each $(x,y)$, $h_\mathrm{min}(x,y)$ and $h_\mathrm{max}(x,y)$,  respectively. The interface is multivalued wherever the
difference $dh(x,y) \equiv h_\mathrm{max}(x,y) - h_\mathrm{min}(x,y)$ is nonzero. 

Looking at Fig. \ref{fig:Method0} one sees that $dh(x,y)$ can be nonzero where there is a vertical cliff or a true overhang with unflipped spins below.
If $N(x,y)$ is the number of interface spins at $(x,y)$, then there will be a cliff with no overhangs where $N(x,y)=dh(x,y)+1$.
The total number of unflipped spins that are part of one or more overhangs
at $(x,y)$ is $\Delta z(x,y)= dh(x,y)-N(x,y)+1$. 

\begin{figure}
	\centering
	\includegraphics[width=0.45\textwidth]{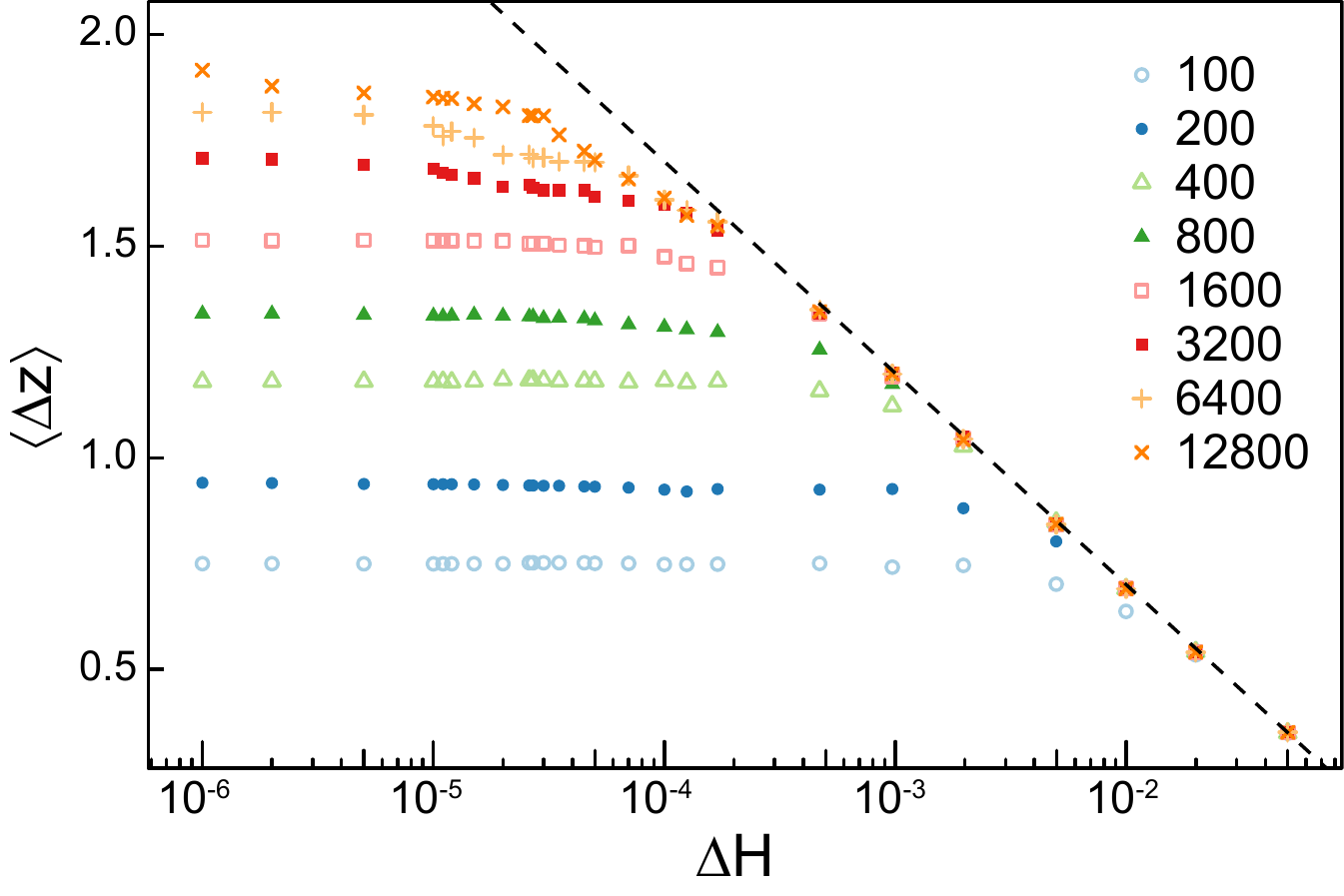}
                                               
	\caption{The average overhang height is calculated at different values of the external fields for the indicated $L$. A line of logarithmic growth is included for comparison.}
	\label{fig:Ap1}
              
\end{figure}

Close to the critical field, approximately an eighth of the projected area of the interface consists of overhangs for $\Delta = 1.7$ as seen in Fig. \ref{fig:Ap0} in Appendix C. This suggests that they could impact the scaling of interface roughness.
However, the average of the total height in overhangs at a given position, $\langle \Delta z \rangle$, grows slowly.
As seen in Fig. \ref{fig:Ap1}, $\langle \Delta z \rangle$ appears to diverge logarithmically as $H \rightarrow H_c$ before saturating at a value that increases roughly logarithmically with $L$.
Reference \cite{Koiller2000} found a similar slow growth in $dh$, which is always greater than $\Delta z$.
Because of the slow growth, the ratios $\langle \Delta z \rangle /L^\chi$ and $\langle dh \rangle/L^\chi$ go to zero as $L \rightarrow \infty$ and $H \rightarrow H_c$ when $\Delta < \Delta_c$.

\begin{figure}
	\centering
	\includegraphics[width=0.45\textwidth]{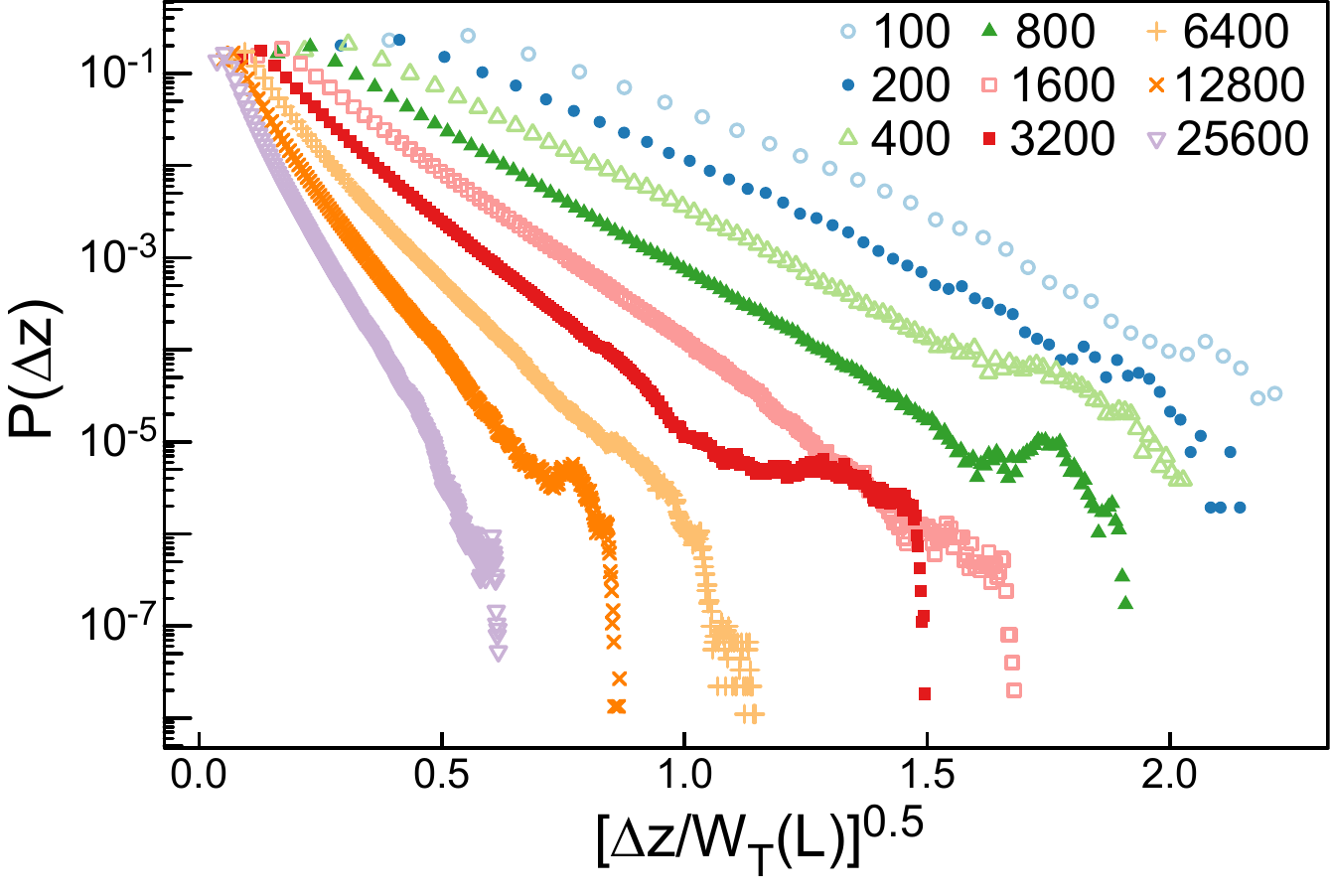}
                                              
	\caption{The distribution $P(\Delta z)$ as a function of $(\Delta z/ W_T(L))^{0.5}$ for the indicated $L$. The data are sampled at $\Delta H = 10^{-6}$. $W_T(L)$ is approximated as $0.13 L^\chi$ for $\chi = 0.85$.}
	\label{fig:Ap2}
             
\end{figure}

Next we consider the probability distribution of individual values of $\Delta z$, $P(\Delta z)$. We find the distribution is well approximated by a stretched exponential with an exponent near 0.5 as shown in Fig. \ref{fig:Ap2}.
Log-linear plots of $P(\Delta z)$ versus $\Delta z^{1/2}$ in Fig. \ref{fig:Ap2} at $\Delta H = 10^{-6}$ follow straight lines until the statistical errors become too large.
To reveal the scaling of overhangs with $L$, $\Delta z^{1/2}$ is normalized by a fit to $W_T(L)$ from Fig. \ref{fig:SP2}, $W_T(L) \approx 0.13 L^\chi$ with $\chi = 0.85$.

Because successive lines in Fig. \ref{fig:Ap2} shift to the left with increasing $L$, overhangs shrink relative to the total rms interface roughness $W_T(L)$ as $L$ increases.
The distribution of fluctuations in the width of the interface from the mean is roughly Gaussian, suggesting the largest overhangs are comparable to the maximum local fluctuations in the height for small $L$. In comparison, at large $L$ the largest overhangs are only a fraction of the rms roughness and much less than the maximum fluctuations in height.
We therefore conclude that overhangs can lead to significant finite-size effects in small systems but are an irrelevant contribution to the surface morphology in the thermodynamic limit. However, overhangs may still be relevant in growth due to their role in overcoming extreme pinning sites.

\section{Conclusion}

Finite-size scaling studies of systems with linear dimensions from 100 to 25600 spins were used to determine critical behavior at the onset of domain wall motion in the 3D RFIM.
Most interface growth models force the interface to be a single-valued function and fix the mean direction of growth. In contrast, an interface in the RFIM can move in any direction, and the driving force is always perpendicular to the local surface.
Nonetheless, the interface breaks symmetry and locks in to a specific growth direction when the rms random field is small enough, $\Delta < \Delta_c \approx 2.5$.
Results are presented for $\Delta =1.7$, but similar scaling
was observed for $\Delta =1.0$, $1.5$, $2$, and $2.1$.
Critical exponents are summarized in Table \ref{table:1}.

\begin{center}
\begin{table}
\begin{tabular}{ |c|c|c|c|c| }
 \hline
  & Values & Prior RFIM & QEW & Predictions \\ 
 \hline
 $\nu_\|$     & 0.79(2)    & 0.75(5) \cite{Ji1992}& 0.80(5) \cite{Leschhorn1993} & 0.77 \cite{Chauve2001} \\  
             &             &  0.77(4) \cite{Roters1999}&  & \\  
             &             & 0.75(2) \cite{Koiller2000}&  & \\  

 $\nu_\perp$  & 0.67(2)    &    & & \\  
 $\alpha$     & 2.84(2)    &    & & \\  
 $\tau$       & 1.280(5)   & 1.28(5) \cite{Ji1992} & 1.30(2) \cite{Rosso2009} & \\
        &    &  & 1.25(2) \cite{LeDoussal2009} & \\
 
 $\tau_\|$    & 1.79(1)   &  & & \\
 $\tau_\perp$ & 1.94(2)   &  & & \\
 $\phi$       & 1.64(4)    & 1.71(11) \cite{Ji1992}  & & \\  
 $\zeta$      & $\ge 0.75$ & 0.67(2) \cite{Ji1992}&  0.75(2) \cite{Leschhorn1993} & 0.86 \cite{Chauve2001} \\  
        &  & 2/3 \cite{Koiller2000} &  0.753(2) \cite{Rosso2003}&  \\  

 $\chi$       & 0.85(1)    &  & & \\
 \hline 
\end{tabular}
\caption{Summary of critical exponents found here for the RFIM and prior results for the RFIM and QEW equation with corresponding references. Prior studies of the RFIM were consistent with $\nu_\perp=\nu_\|$ and $\chi=1$. Scaling relations involving these exponents are found in Eqs. \eqref{eqn:phi}, \eqref{eqn:txz}, \eqref{eqn:ts}, and \eqref{eq:scale2}. Uncertainties in the last digit are indicated in parentheses.}
\label{table:1}
\end{table}
\end{center}

In an infinite system there is a transition at $H_c$ from motion through unstable jumps between stable states to steady motion at a nonzero velocity.
In a finite system the transition occurs over a finite range of fields.
Near $H_c$ there is a growing probability that avalanches may span the system and even advance the entire system (fully spanning avalanches).
Finite-size scaling of the fraction of volume invaded by spanning and fully spanning avalanches was used to determine $H_c$ and the in-plane correlation length exponent $\nu_\|$ (Fig. \ref{fig:V0}).
Past studies used either the fraction of sites invaded in a cubic system \cite{Ji1992} or the probability of spanning a cubic system \cite{Koiller2000}.
This overestimates $H_c$ because growth is anisotropic and the typical height of the interface at $H_c$ is only of order $L^\chi$.
The correlation length exponent is also affected.

As $H$ approaches $H_c$ from below, the mean volume of avalanches grows as $\langle S \rangle \sim \Delta H ^{-\phi}$ until it saturates due to the finite system size.
The value of $\phi$ and an independent measure of $\nu_\|$ are obtained
by scaling results for different $L$
[Fig. \ref{fig:V1} and Eq. \eqref{eqn:ave}].
At $H_c$ the probability distribution of $S$ decreases as $S^{-\tau}$ up to a maximum size that scales as $L^\alpha$ (Fig. \ref{fig:V2}).
The values of $\alpha$ and $\tau$ were determined by scaling the distributions for different $L$.
Independently determined exponents agreed with the scaling relation given in Eq. \eqref{eqn:phi}. 

The mean height and width of avalanches and their distributions must obey analogous scaling relations. Finite-size scaling collapses in Sec. IIID test these relations and reveal a clear anisotropy in growth (Fig. \ref{fig:M2}).
The height of avalanches $\ell_\perp$ diverges as $H \rightarrow H_c$ with an exponent $\nu_\perp$ that differs from $\nu_\|$. The height and width of individual avalanches are related
by $\ell_\perp \sim \ell_\|^\chi$ with $\chi=\nu_\perp/\nu_\|$ (Fig. \ref{fig:M1}).
The divergence of the mean height of the interface is consistent with the growth in the size of the largest avalanche: $\nu_\perp =1-\phi$ [Eq. \eqref{eq:scale2}].

Table \ref{table:1} contrasts results obtained here with past studies of the RFIM and related models.
References \cite{Ji1992} and \cite{Koiller2000} assumed $\chi=1$. This leads to a reduced set of scaling relations that were consistent with their exponents.
Note that their values of $\nu_\|$, $\tau$, and $\phi$ are consistent with our results but have much larger error bars because of the smaller system sizes available.
Slightly larger systems in Ref. \cite{Nolle1996} gave an indication that 
$\chi$ was less than unity, but could not rule out $\chi=1$. 

The largest difference from past work on the RFIM is the value of $\zeta$.
References \cite{Ji1992, Koiller2000} considered the scaling of roughness with $\ell$ at a given $L$
and found results were consistent with the mean-field value of 2/3.
As seen in Sec. IIIF, this measure is strongly affected by system size.
The slope on log-log plots rises continuously as $H$ goes to $H_c$ and as $L$ increases.
Results for $L \sim 1000$ are consistent with $\zeta \approx 2/3$, but
values up to 0.75 are observed for $L =12800$ (Fig. \ref{fig:w0}).
These changes appear to be related to overhangs that lead to growing roughness at small scales as $L$ increases.
The results in Sec. IIIH support the conclusion that these changes become irrelevant
at the critical point.
We find that the total rms roughness is not significantly affected by overhangs and scales as $L^\chi$ with $\chi=0.85 \pm 0.02$ for all $L \ge 100$ (Fig. \ref{fig:SP2}).

Table \ref{table:1} also includes results for the evolution of single-valued interfaces governed by the QEW equation.  
Estimates of the avalanche distribution exponent $\tau$ \cite{Rosso2009, LeDoussal2009} are consistent with the value measured in Sec. IIIC for the RFIM. 
The roughness exponent $\zeta$ found in the QEW equation is consistent with our lower bound for $\zeta$ although it is distinct from $\chi$ \cite{Rosso2003}. 
Interestingly, results from two-loop functional renormalization group analysis indicate $\zeta = 0.86$ \cite{Chauve2001}. This prediction for $\zeta$ is even closer to the exponent $\chi$ identified in this paper. Finally, scaling relation results from simulations \cite{Leschhorn1993} of $\nu_\| = 0.80\pm 0.05$ and the two-loop renormalization group result \cite{Chauve2001} of $\nu_\| = 0.77$ cannot be distinguished from our measurement of $\nu_\|$ for the RFIM.

Comparing numerically measured exponents for the RFIM and QEW equation, one cannot conclusively determine whether they are distinct. However, our measure of $\zeta$ is a lower bound which we anticipate will approach $\chi = 0.85$ with increasing $L$, while simulations of the QEW give the smaller value of $\zeta = 0.75$ \cite{Leschhorn1993}. This difference suggests the RFIM resides in a different universality class than the QEW equation. Note that the QEW exponents are believed \cite{Nattermann1992, Narayan1993} to obey an additional relation $\nu_\| = 1/(2-\zeta)$. Our exponents are not consistent with this relation and others that follow from it if $\zeta = \chi$. Furthermore, although the morphology of overhangs becomes irrelevant in the thermodynamic limit, the ability of a fully $d$ dimensional interface to grow laterally is still important and fundamentally changes the system's response to extreme pinning sites. We find more than 10\% of the projected area consists of overhangs indicating lateral growth is an important mechanism in the propagation of RFIM domain walls. 

The anisotropy of individual avalanches has not yet been measured in the $d=2+1$ QEW equation; however, Rosso \textit{et al.} \cite{Rosso2009} measured the maximum size of avalanches in $d = 1+1$ and found it scales as $\xi^{1+\zeta}$. The anisotropy of avalanches has also been directly studied in single-valued models of directed percolation depinning (DPD) in $d = 2+1$, producing results consistent with $\chi = \nu_\perp /\nu_\| = \zeta$ where $\zeta = 0.58 \pm 0.03$ \cite{Amaral1995}. 
It is interesting that $\chi = \zeta$ in DPD although it is important to note that DPD resides in a distinct universality class described by the quenched Kardar-Parisi-Zhang equation \cite{Kardar1986, Amaral1994}. 

The studies presented here show that finite-size effects remain important until very large system sizes and small $\Delta H$. Given recent conclusions about the importance of rare events in the QEW model \cite{Cao2018}, it would be interesting to extend past QEW studies to the much larger sizes studied here.
Further studies on the RFIM and QEW models are also needed to clarify the relation between $\chi$ and $\zeta$.
This work clearly identifies an anisotropy exponent $\chi=0.85$ in several independent measures that have not been applied to the QEW.
While the roughness exponent measured for individual interfaces approaches $\chi$, it remains significantly below $\chi$ even for $L=12800$.
An important topic for future work will be to confirm that $\zeta$ approaches $\chi$ as predicted by current scaling theories or show that $\zeta$ remains distinct from $\chi$, implying new theories are needed.

\begin{acknowledgments}
The authors thank Karin Dahmen and Alberto Rosso for useful conversations. This material is based upon work supported by the National Science Foundation under Grant No. DMR-1411144. MOR acknowledges support from the Simons Foundation. Calculations were performed on Bluecrab at the Maryland Advanced Research Computing Center.
\end{acknowledgments}

\appendix
\section{Volume invaded}

In Sec. IIIB, the scaling of the average volume of an avalanche $\langle S \rangle$ was determined. Here the analysis is extended to develop a scaling relation for the divergence of the total integrated volume.
Over a small increase in external field from $H$ to $H + d H$, the interface will advance a volume $d V$:
\begin{equation}
d V \sim \langle S \rangle_A A(H,L) R(H,L) d H\ \ , 
\label{eqn:app1}
\end{equation}
where $A$ is the number of spins on the external interface and the nucleation
rate $R$ is the number of avalanches nucleated per spin per change in field. Here, $\langle S\rangle_A$ indicates an average over all avalanches including spanning avalanches. Since the largest avalanches dominate $\langle S \rangle_A$ for $\tau<2$, spanning avalanches contribute most to $d V$. This explains why $F_s$ is near unity close to $H_c$ (Fig. \ref{fig:V0}). We begin by studying how $A$ and $R$ evolve with increasing $H$ and $L$.

The area $A$ is defined as the number of flipped spins that are on the external interface and adjacent to unflipped spins. One could also count the number of unflipped spins adjacent to these flipped spins or the number of bonds between flipped and unflipped spins. These measures differ by less than 0.1\% for all $H$ and $L$ and thus give the same scaling behavior.

The area of the interface is initially equal to $L^2$. As the interface advances and roughens, $A$ increases. Even a single-valued rough interface defined on the cubic lattice will have $A > L^2$ because of discrete steps in height on the lattice. If the interface steps up by $n$ sites, there will be $n$ spins on the interface at the same $x,y$. Overhangs, as discussed in Sec. IIIH and Appendix C, produce a further increase in $A$ because there may be multiple horizontal interfaces at each $x,y$.

\begin{figure}
	\includegraphics[width=0.45\textwidth]{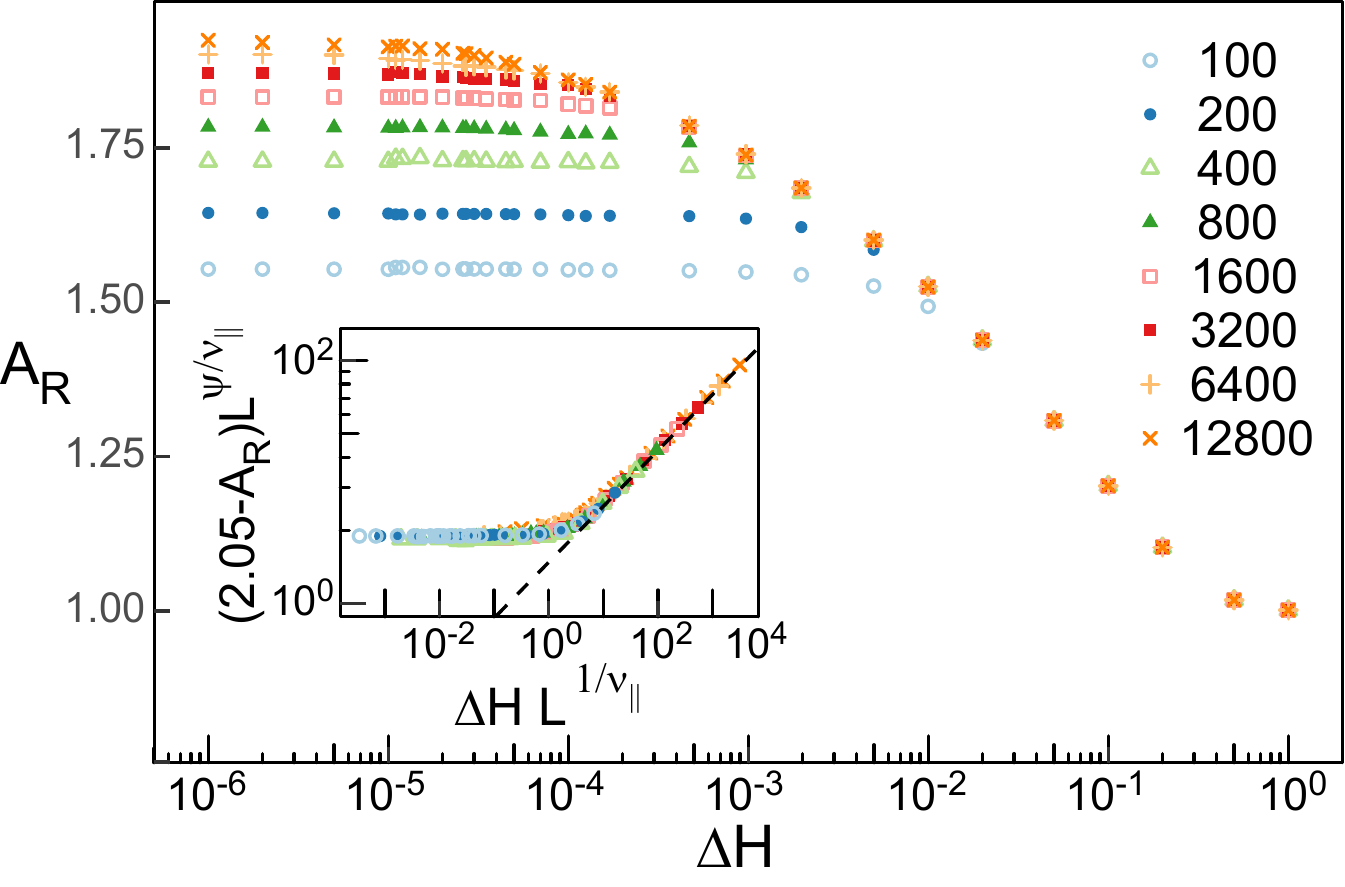}
	\caption{The surface area of the external interface normalized by $L^2$ is shown as a function of external field for $L$ given in the legend. Inset: Collapse of data in main panel for $\Delta H < 10^{-2}$ using Eq. \eqref{eq:AR} for $A_{R,\mathrm{Lim}}=2.05$, $\psi = 0.23$ and $\nu_\| = 0.79$. Data were generated using the alternative growth protocol where spins are flipped until a stable interface is reached at a fixed value of the external field.}
	\label{fig:AA0}
\end{figure}

To remove the trivial dependence of area on $L^2$, we define the relative area $A_R(H,L) \equiv A(H,L)/L^2$. Figure \ref{fig:AA0} shows how $A_R$ grows as $H$ approaches $H_c$. For each $L$, the value of $A_R$ saturates as $\Delta H$ decreases. The onset of saturation occurs at the same $\Delta H$ as other quantities discussed in the main text, and is associated with $\xi_\|$ reaching  $L$.
In contrast to other quantities, the limiting value of $A_R$ remains finite.
Since Fig. \ref{fig:AA0} is a linear-log plot, one can see that 
$A_R$ grows less than logarithmically with increasing $L$.
The inset of Fig. \ref{fig:AA0} shows that the data are consistent with convergence to a finite limiting value $A_{R,\mathrm{Lim}}=2.05$ as $L\sim \infty$.
Data for all $L$ and $\Delta H < 10^{-2}$ are collapsed by assuming a power-law approach to $A_{R,\mathrm{Lim}}$ with an exponent $\psi$:
\begin{equation}
A_{R,\mathrm{Lim}} - A_R(H,L) \sim L^{-\psi/\nu_\|} f_A\left( \Delta H L^{1/\nu_\|} \right) \ \ ,
\label{eq:AR}
\end{equation}
where $f_A(x)$ is a new scaling function that saturates for $x \ll 1$ and
scales as $f_A(x) \sim x^{\psi}$ for $x \gg 1$.
The quality of the collapse is consistent with $\psi = 0.23 \pm 0.05$ and $A_{R,\mathrm{Lim}} = 2.05 \pm 0.05$. Due to the dependence on many parameters, it is difficult to get more accurate estimates of these values.

The value of $A_{R, \mathrm{Lim}}$ increases with the strength of the noise $\Delta$. For $\Delta = 2.1$, we find $A_{R, \mathrm{Lim}} = 3.52 \pm 0.05$ with the same value of $\psi$ within our errorbars. In the self-similar regime ($\Delta > \Delta_c$), the surface area of the invaded volume will scale at least as rapidly as $L^{D_f}$, where $D_f>2$ is the fractal dimension. Therefore, we expect $A_{R,\mathrm{Lim}}$ to diverge as $\Delta$ approaches $\Delta_c$, but we do not study this transition here.

\begin{figure}
	\includegraphics[width=0.45\textwidth]{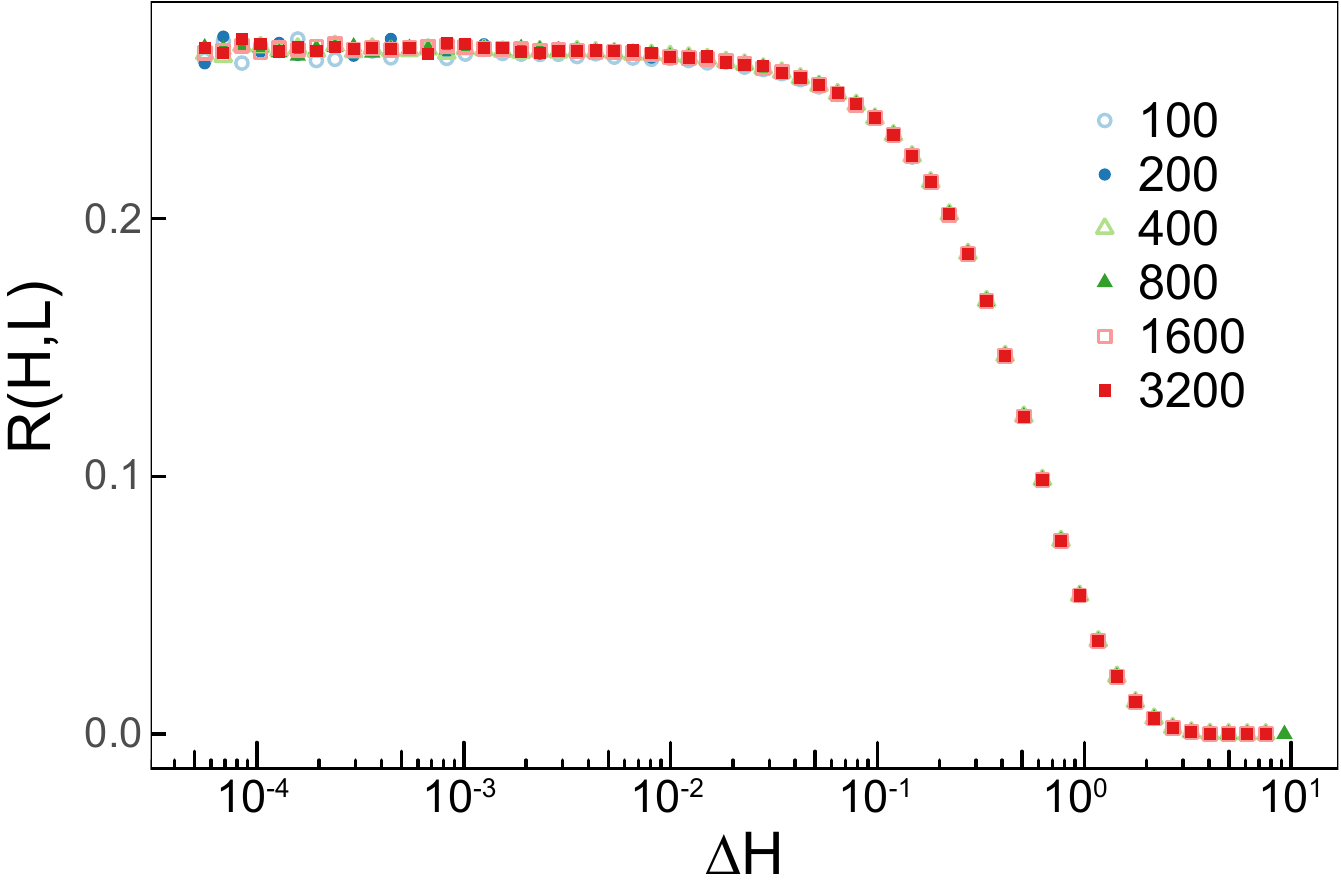}
	\caption{The rate of avalanche nucleation $R(H,L)$ over logarithmically spaced intervals of $H$ for values of $L$ indicated in the legend. }
	\label{fig:AA1}
\end{figure}

We now turn our focus to the rate $R(H,L)$ at which avalanches nucleate per spin per increment of external field. The rate $R$ is calculated by tallying the number of avalanches, both spanning and nonspanning, nucleated over an interval of field and then dividing the total by the duration of the interval and the surface area. Intervals are evenly spaced on an axis of $\log \Delta H$. As seen in Fig. \ref{fig:AA1}, $R$ does not depend on $L$ and becomes independent of $\Delta H$ for $\Delta H < 10^{-2}$. This is part of the evidence used in the main text to determine that the critical region is limited to $\Delta H < 10^{-2}$. Figure \ref{fig:AA1} confirms that sufficiently close to the critical point the nucleation rate is constant and extensive with the surface area. As noted in the main text, this is expected for interface motion but not for sheared systems \cite{Lin2015, Lin2014a, Salerno2012, Salerno2013}.

Equipped with these results, we now derive an expression for the total volume invaded in an infinite system where $\langle S \rangle_A = \langle S \rangle$. Equation \eqref{eqn:app1} can be rewritten as:
\begin{equation}
d V \sim \langle S \rangle L^2 A_R (H,L) R(H,L) d H \ \ .
\label{eqn:app2}
\end{equation}
From Eq. \eqref{eqn:ave2}, $\langle S \rangle \sim \Delta H^{-\phi}$ diverges as $H$ approaches $H_c$. For $\Delta H$ small enough that $A_R$ and $R$ are approximately constant, Eq. \eqref{eqn:app2} can be integrated to yield
Eq. \eqref{eq:Vtot}.

\begin{figure}
	\includegraphics[width=0.45\textwidth]{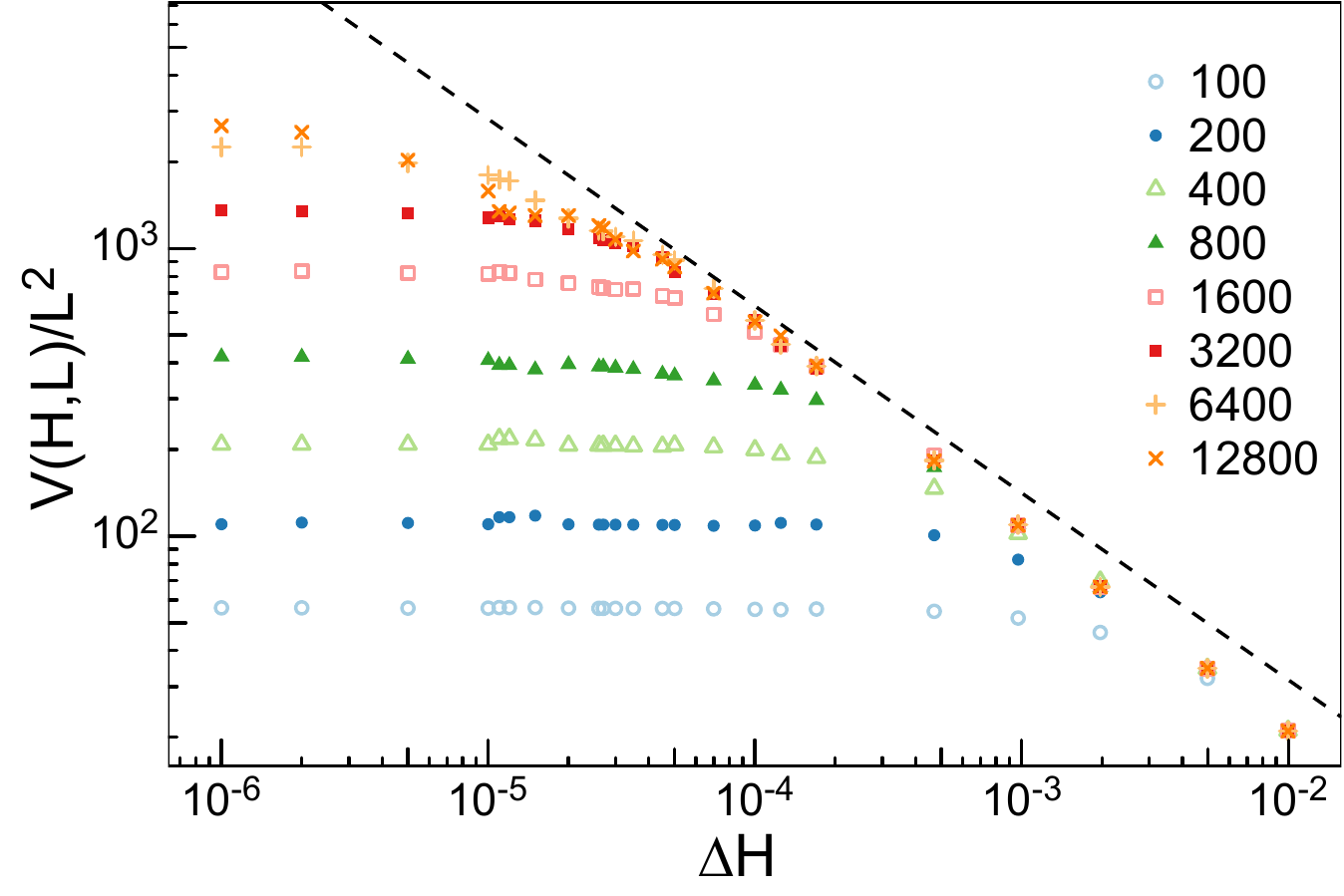}
	\caption{The total volume invaded normalized by $L^2$ is calculated as a function of $H$ for values of $L$ indicated in the legend. A power law with exponent $1-\phi = -0.64$ is overlaid for comparison and follows the data for about a decade.}
	\label{fig:AA3}
\end{figure}

Figure \ref{fig:AA3} shows $V/L^2$ as a function of $H$ and $L$.
A dashed line indicates the expected power-law divergence using the value of $\phi=1.64$ from Sec. IIIB.
The data appears to follow the expected scaling for about a decade from $5 \times 10^{-5}$ to $5 \times 10^{-4}$. Finite-size effects set in at smaller $\Delta H$. As shown above, the variation in $A_R$ remains significant down to $\Delta H \sim 5 \times 10^{-4}$.

\section{Distribution of avalanches including spanning events}

\begin{figure}
	\centering
	\includegraphics[width=0.45\textwidth]{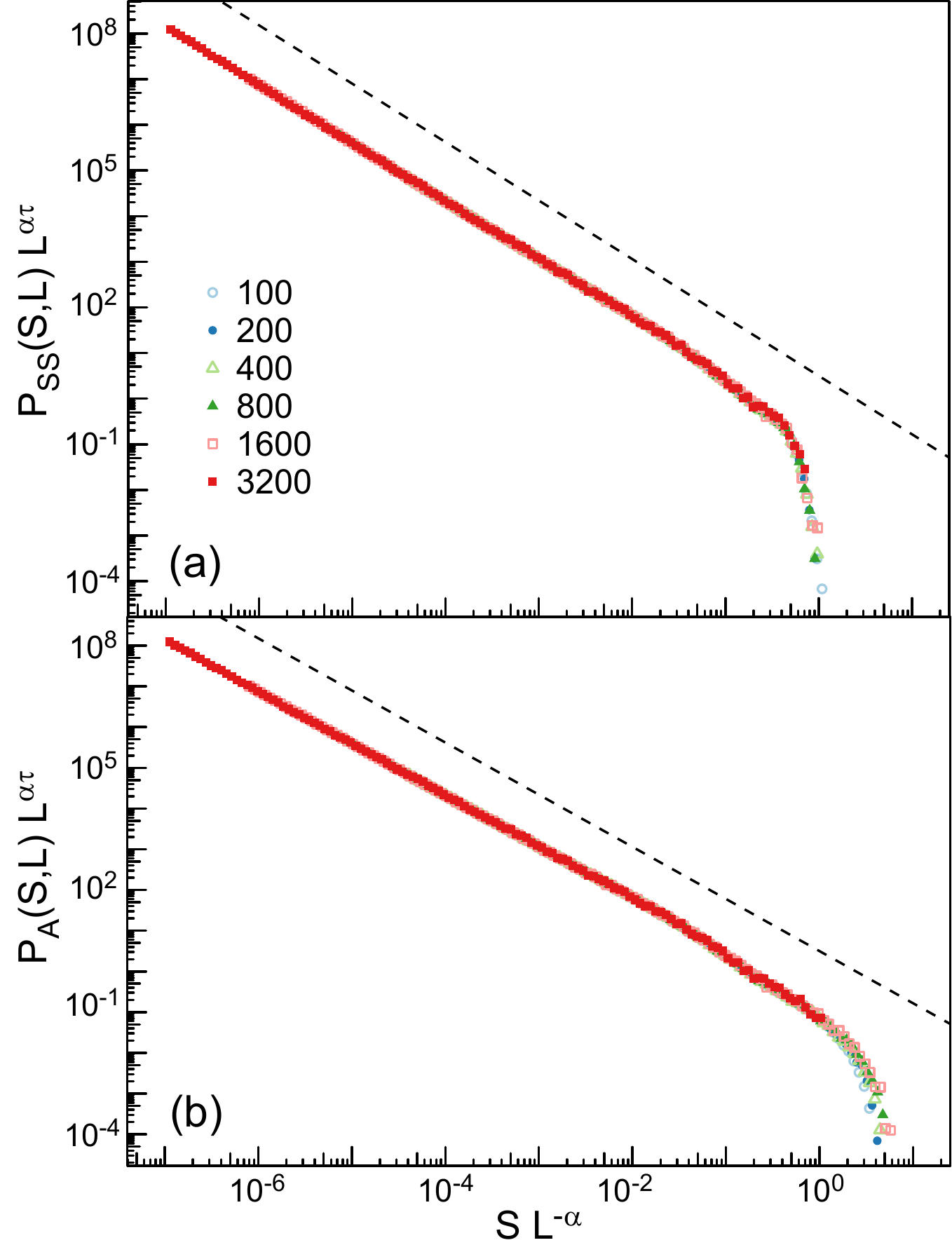}
                                                   
	\caption{The probability distribution of avalanche volumes is calculated including (a) nonspanning and semispanning avalanches and (b) all avalanches for values of $L$ given in the legend. Data are rescaled using the scaling relation in Eq. (9) with exponents $\tau = 1.28$ and $\alpha = 2.84$. The dashed lines show a power-law decay with $\tau=1.28$. }
	\label{fig:AB}
             
\end{figure}

In this appendix, we expand on the discussion of semi- and fully spanning avalanches in Sec. IIIE. Semispanning avalanches were argued to behave more like nonspanning avalanches while fully spanning avalanches were argued to be more representative of depinning motion. To confirm this finding, we extend the definition of the the probability distribution of avalanche volumes defined in Sec. IIIC,  $P(S,L)$, to include semispanning avalanches, $P_{SS}(S, L)$, as well as all avalanches, $P_{A}(S,L)$. These distributions are calculated using the same method used in Fig. \ref{fig:V2}a. In Fig. \ref{fig:AB},  (a) $P_{SS}(S, L)$ and (b) $P_A(S, L)$ are scaled using the scaling ansatz in Eq. \eqref{eqn:ps2}. 

These collapsed curves both resemble those of $P(S, L)$ in Fig. \ref{fig:V2}(b) for $S/L^\alpha \lesssim 10^{-2}$. Above this scale, the distributions deviate due to the inclusion of spanning avalanches.  The upper cutoff for $P_{SS}(S, L)$ is seen to collapse in  Fig. \ref{fig:AB}(a). Therefore, the maximum volume of semispanning avalanches grows as $L^\alpha$. This is in agreement with the behavior seen in Fig. \ref{fig:M3} where semispanning avalanches scale in the same manner as nonspanning avalanches.

However, the maximum volume of fully spanning avalanches no longer scales as $L^\alpha$. As seen in Fig. \ref{fig:AB}(b), $P_{A}(S, L)$ has two drops at large $S$. The first scales as $L^\alpha$ and reflects the limiting size of semispanning avalanches. The second drop occurs at a threshold scaling as $L^3$ and is due to fully spanning avalanches. Avalanches are ultimately limited by the finite volume of the box that scales as $L^3$. As a fully spanning avalanche has a width of $L$, this scaling implies the maximum height scales in proportion to $L$, the maximum height of the box. This agrees with the scaling seen in Fig. \ref{fig:M3} and further suggests fully spanning avalanches are more closely related to behavior above the depinning transition.


\section{Additional overhang statistics}

In Sec. IIIH, overhangs were defined as multivalued regions of the projected interface and their heights were characterized. In this appendix, we provide additional information on the number of overhangs and spatial clustering of overhangs.

\begin{figure}
	\centering
	\includegraphics[width=0.45\textwidth]{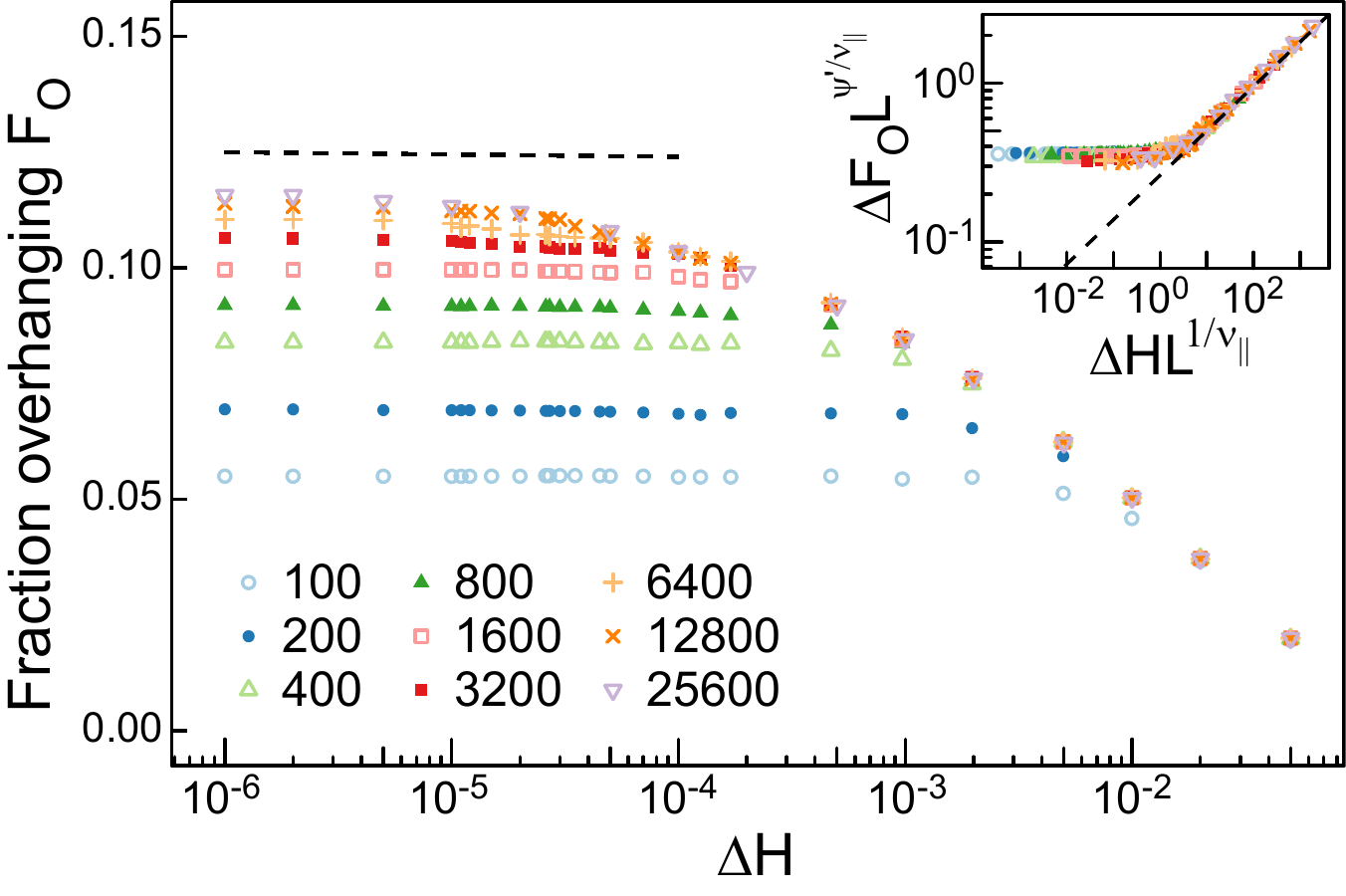}
                                                   
	\caption{The percentage of the projected area for which there are overhangs on the interface is calculated as a function of distance from the critical field at the $L$ indicated in the legend. A horizontal dashed line indicates $0.124$. Inset: The data in the main panel are collapsed using a similar finite-size scaling ansatz to Eq. \eqref{eq:AR} where $\Delta F_O = 0.124 - F_\mathrm{O}$ and $\psi = 0.3$. A dashed line with slope 0.3 is overlaid on the data.}
	\label{fig:Ap0}
             
\end{figure}

The fraction of the projected interface that contains an overhang, $F_\mathrm{O}$, is just the fraction of $(x,y)$ where $\Delta z$ is nonzero.
Figure \ref{fig:Ap0} shows how $F_O$ evolves with $\Delta H$ and $L$. Initially, the interface is flat and $F_O$ is zero for all $L$. As the system approaches the critical point, $F_O$ grows for all $L$ before saturating at a field $\Delta H$ that decreases with increasing $L$. The saturating percentage rises more slowly than logarithmically with $L$ and appears to approach an asymptotic limit between 12 and 13\% as $L \rightarrow \infty$. Assuming that the difference $\Delta F_O$ from the asymptotic value decays as $\Delta H^{-\psi'}$, one can derive a scaling relation analogous to Eq. \eqref{eq:AR}. As shown in the inset of Fig. \ref{fig:Ap0}, the data can be collapsed fairly well with $\psi' = 0.3 \pm 0.05$ and a limiting fraction of $0.124\pm 0.005$.
Note that the fraction of the surface where cliffs occur is roughly twice $F_O$, and that both fractions increase as $\Delta$ rises towards $\Delta_c$.

\begin{figure}
	\centering
	\includegraphics[width=0.45\textwidth]{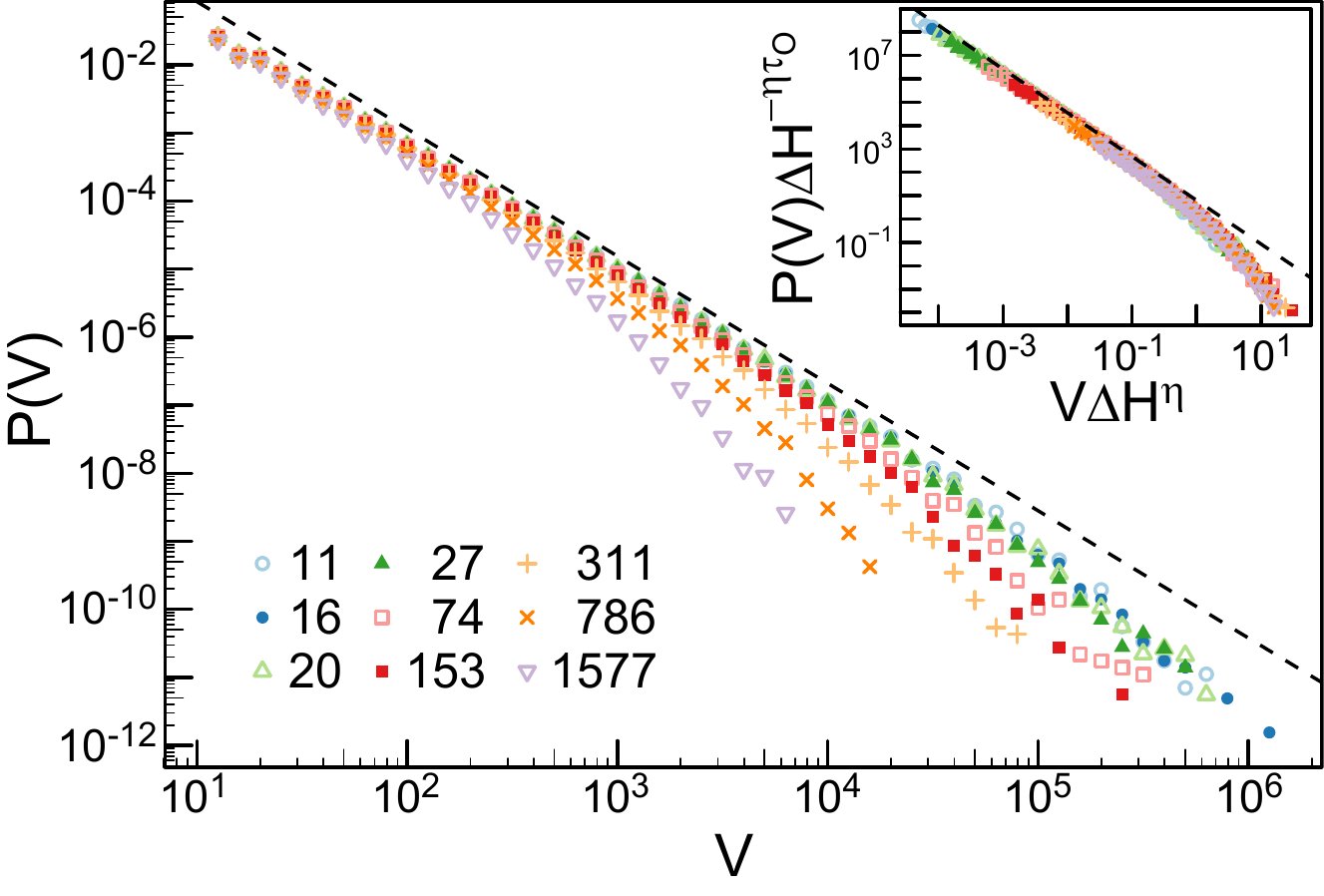}
                                               
	\caption{The probability distribution of the volume $V$ of aggregated overhangs is calculated at values of $\Delta H L^{1/\nu_\|}$ indicated in the legend for $L = 12800$. Note that $\Delta H < 0.01$ for all curves. A dashed line indicates a power law of exponent 1.87. Inset: The data in the primary panel are collapsed using Eq. \eqref{eqn:apoh} and values of $\eta = 1.3$ and $\tau_{O} = 1.87$.}
	\label{fig:Ap3}
             
\end{figure}

Overhangs are not isolated features and one expects there to be lateral correlations. To account for lateral structure, we clustered adjacent $(x, y)$ locations where $\Delta z > 0$ into aggregated overhangs and calculated the total volume $V$ of each aggregated overhang. The volume is simply defined as the sum of all the clustered values of $\Delta z$. The probability distribution of $V$ decays as a power of $V$ with an exponent consistent with $\tau_{O} \sim 1.87 \pm 0.05$ as seen in Fig. \ref{fig:Ap3}. This power law extends to an upper cutoff $V_\mathrm{max}$ that increases as $H \rightarrow H_c$. Assuming $V_\mathrm{max} \sim \Delta H^{-\eta}$ with $\eta$ a new exponent, we propose the following scaling ansatz:
\begin{equation}
P(V) \sim \Delta H^{\eta \tau_{O}} g_{O}(V \Delta H^\eta) \  \ ,
\label{eqn:apoh}
\end{equation}
where $g_{O}(x)$ is a universal scaling function which scales as $x^{-\tau_{O}}$ for $x \ll 1$ and rapidly decays to zero for $x \gg 1$. This relation will hold only before the onset of finite-size effects at $\Delta H L^{1/\nu_\|} \approx 10$. Using this relation, the data in Fig. \ref{fig:Ap3} are collapsed in the inset. Based on the sensitivity of the collapse, we estimate $\eta \sim 1.3 \pm 0.1$ and $\tau_{O} \sim 1.87 \pm 0.05$. The relation of these exponents to others is currently unknown.

The exponent $\tau_{O} <2$, and so the arguments used in Sec. IIIC imply that the volume of the largest overhangs will dominate the average volume.
Figure \ref{fig:Ap3} implies that the maximum volume diverges
as $H \rightarrow H_c$, so the characteristic volume of an aggregated overhang will also diverge. However, this divergence is considerably slower than the divergence of the volume of the largest avalanche, which scales as $\Delta H^{-\nu_\| \alpha} \sim \Delta H^{-2.25}$. Thus as with other results in Sec. IIIH, the nontrivial scaling of overhangs may lead to interesting finite-size effects but becomes irrelevant at the critical field in infinite systems.

%

\end{document}